\documentclass[usenatbib]{basi}
\usepackage{graphicx}
\usepackage{times,amssymb}
\usepackage[fleqn,tbtags]{amsmath}

\usepackage[british]{babel}
\usepackage[varg]{txfonts}
\usepackage{rotating}
\usepackage{dcolumn}

\def\rnl{r_\mathrm{nl}}

\begin{document}

\title[Halo formation rates]{Formation rates of dark matter haloes} 

\author[S. Mitra et al.]{Sourav Mitra$^1$\thanks{email: \texttt{smitra@hri.res.in}}, 
Girish Kulkarni$^{1,2}$\thanks{email: \texttt{girish@mpia-hd.mpg.de}}, 
J. S. Bagla$^{1,3}$\thanks{email: \texttt{jasjeet@iisermohali.ac.in}} 
and Jaswant K. Yadav$^{4}$\thanks{email: \texttt{jaswant@kias.re.kr}}\\
$^1$Harish-Chandra Research Institute, Chhatnag Road, Jhunsi, Allahabad 211019, India\\ 
$^2$Max-Planck Institut f\"ur Astronomie, K\"onigstuhl 17, D-69117 Heidelberg, Germany\\
$^3$Indian Institute of Science Education and Research (Mohali), Sector 81,\\ S
A S Nagar, Manauli PO, Mohali 140306, Punjab, India \\
$^4$Korea Institute for Advanced Study, Hoegiro 87, Dongdaemun-gu, Seoul 130722, South Korea} 

\pubyear{2011}
\volume{39}
\pagerange{\pageref{firstpage}--\pageref{lastpage}}

\date{Received 2011 August 08; accepted 2011 November 21}

\maketitle

\label{firstpage}

\begin{abstract}
We derive an estimate of the rate of formation of dark matter haloes per unit
volume as a function of the halo mass and redshift of formation.  
Analytical estimates of the number density of dark matter haloes are useful in
modeling several cosmological phenomena.   
We use the excursion set formalism for computing the formation rate of dark
matter haloes. 
We use an approach that allows us to differentiate between major and minor
mergers, as this is a pertinent issue for semi-analytic models of galaxy
formation. 
We compute the formation rate for the Press-Schechter and the Sheth-Tormen mass
function. 
We show that the formation rate computed in this manner is positive at all
scales. 
We comment on the Sasaki formalism where negative halo formation rates are
obtained. 
Our estimates compare very well with N-body simulations for a variety of
models. 
We also discuss the halo survival probability and the formation redshift
distributions using our method.
\end{abstract}

\begin{keywords}
cosmology: large scale structure of the Universe -- galaxies: formation --
galaxies: evolution -- galaxies: haloes 
\end{keywords}

\section{Introduction}

Gravitational amplification of density perturbations is thought to be
responsible for formation of large scale structures in the Universe
\citep{1980lssu.book.....P, 1989RvMP...61..185S, 1999coph.book.....P,
  2002tagc.book.....P, 2002PhR...367....1B}.  Much of the matter is
the so called dark matter that is believed to be weakly interacting
and non-relativistic \citep{1987ARA&A..25..425T, 2009ApJS..180..330K}.
Dark matter responds mainly to gravitational forces, and by virtue of
a larger density than baryonic matter, assembly of matter into haloes
and large scale structure is primarily driven by gravitational
instability of initial perturbations in dark matter.  Galaxies are
believed to form when gas in highly over-dense haloes cools and
collapses to form stars in significant numbers
\citep{1953ApJ...118..513H, 1977MNRAS.179..541R, 1977ApJ...211..638S,
  1977ApJ...215..483B}.  Thus the hierarchical formation of dark
matter haloes is the key driver that leads to formation and evolution
of galaxies and clusters of galaxies.

The halo mass function describes the comoving number density of dark
matter haloes as a function of mass and redshift in a given cosmology.
It is possible to develop the theory of mass functions in a manner
that makes no reference to the details of the cosmological model or
the power spectrum of fluctuations.  That is, we expect the mass
function to take a universal form, when scaled appropriately.  Simple
theoretical arguments have been used to obtain this universal
functional form of the mass function \citep{1974ApJ...187..425P,
  1991ApJ...379..440B, 2001MNRAS.323....1S}.
\citet{1991ApJ...379..440B}, and, \citet{2001MNRAS.323....1S} used the
excursion set theory to derive the mass function.  Much work has also
been done to determine the extent to which this form is consistent
with results from N-body simulations \citep{2001MNRAS.321..372J,
  2002ApJS..143..241W, 2003MNRAS.346..565R, 2006ApJ...646..881W,
  2007MNRAS.374....2R, 2007ApJ...671.1160L, 2008MNRAS.385.2025C,
  2008ApJ...688..709T} with the conclusion that the agreement is
fairly good.  It is remarkable that a purely local approach provides a
fairly accurate description of the manifestly non-linear and strongly
coupled process of gravitational clustering.  The success of the local
description has been exploited in developing the semi-analytic
theories of galaxy formation \citep{1991ApJ...379...52W,
  1993MNRAS.264..201K, 2000ApJ...534..507C, 2001ApJ...555...92M,
  2007MNRAS.377..285S}.

The Press-Schechter mass function \citep{1974ApJ...187..425P} that is commonly
used in these semi-analytic models assumes spherical collapse of haloes
\citep{1972ApJ...176....1G}. 
The shape of this mass function agrees with numerical results qualitatively,
but there are deviations at a quantitative level \citep{1988MNRAS.235..715E,
  2001MNRAS.321..372J}.   
Improvements to the Press-Schechter mass function have been made to
overcome this limitation.  
In particular, the Sheth-Tormen mass function, which is based on the more
realistic ellipsoidal collapse model \citep{1999MNRAS.308..119S,
  2001MNRAS.323....1S} fits numerical results better. 
Many fitting functions with three or four fitting parameters have been
proposed, these are based on results of simulations of the Lambda-Cold Dark 
Matter ($\Lambda$CDM) model
\citep{2001MNRAS.321..372J, 2003MNRAS.346..565R, 2006ApJ...646..881W,
  2010MNRAS.406.2267F}.  

In the application of the theory of mass functions to the semi-analytic models
for galaxy formation, we often need to know comoving number density of haloes
of a certain age. 
Naturally, this quantity is related to the halo formation rates and the
survival probability.  
While these details are known and well understood for the Press-Schechter mass
function \citep{1974ApJ...187..425P}, the situation is not as clear for other
models of the mass function. 
Furthermore, analytic estimates for the halo formation rate and survival
probability are important in spite of the availability of accurate fitting
functions for these quantities in the $\Lambda$CDM model. 
This is because analytic estimates can be used to study variation in these
quantities with respect to, for instance, the underlying cosmology or the
power spectrum of matter perturbations. 
Studying such variation with the help of simulations is often impractical.
In this work, we focus on the computation of halo formation rates.

Several approaches to calculating halo formation 
rates have been suggested
\citep{1993MNRAS.264..509B, 1994PASJ...46..427S, 1996MNRAS.280..638K}.
In particular, \citet{1994PASJ...46..427S} suggested a very simple
approximation for the formation rate as well as survival probability for
haloes. 
The approximation was suggested for the Press-Schechter mass function, though
it does not use any specific aspect of the form of mass function. 
The series of arguments is as follows:
\begin{itemize}
\item 
Merger and accretion lead to an increase in mass of individual haloes.
Formation of haloes of a given mass from lower mass haloes leads to an increase
in the number density, whereas destruction refers to haloes moving to a higher
mass range.  The net change in number density of haloes in a given interval in
mass is given by the difference between the formation and destruction rate.  
\item 
Given the net rate of change, we can find the formation rate if we know the
destruction rate. 
\item 
A simple but viable expression for the destruction rate is obtained by
assuming that the probability of destruction per unit mass (also known as the
halo destruction efficiency) is independent of mass. 
\item 
This approximate expression for the destruction rate is then used to derive
the formation rate as well as the survival probability.  
\end{itemize}

The resulting formulae have been applied freely to various cosmologies
and power spectra, including the CDM class of power spectra.  The
Sasaki approach has been used in many semi-analytic models of galaxy
formation \citep{2000ApJ...534..507C, 2005MNRAS.361..577C,
  2007MNRAS.377..285S} mainly due to its simplicity.  Attempts have
also been made to generalize the approximation to models of mass
function other than the Press-Schechter mass function
\citep{2009NewA...14..591S}, though it has been found that a simple
extension of the approximation sometimes leads to unphysical results.
In particular, when applied to the Sheth-Tormen mass function, the
Sasaki approach yields negative halo formation rates.

In this paper, we investigate the application of the Sasaki approach
to the Sheth-Tormen mass function.  We test the Sasaki approach by
explicitly computing the halo formation and destruction rates for the
Press-Schechter mass function using the excursion set formalism.  We
then generalize this same method to compute the halo formation rates
for the Sheth-Tormen mass function.  We find that halo formation rates
computed in this manner are always positive.  Finally, we compare our
analytical results with N-body simulations.

A reason for choosing the approach presented in this paper, as compared to
other competing approaches based on the excursion set formalism, is that we
wish to be able to differentiate between major and minor mergers. 
This is an essential requirement in semi-analytical models of galaxy formation
and is not addressed by other approaches for halo formation rate
\citep{1999MNRAS.309..823P, 2000MNRAS.318..273P, 2001MNRAS.327.1313P,
  2007MNRAS.376..977G, 2008MNRAS.391.1729M, 2009MNRAS.397..299M}.

Many previous studies of merger rates using analytical or numerical
techniques are present in the literature.  \citet{2005MNRAS.357..847B}
recognised that the Sasaki approach of calculating halo formation rate
was fundamentally inconsistent.  They showed that a mathematically
consistent halo merger rate should yield current halo abundances when
inserted in the Smoluchowski coagulation equation.  They applied this
technique to obtain merger rates for the Press-Schechter mass
function.  The original formulation of halo merger rates in the
excursion set picture \citep{1994MNRAS.271..676L} was also improved by
\citet{2008MNRAS.388.1792N} and \citet{2010MNRAS.403..984N} to include
the effect of finite merger time interval.  They found that the
resultant merger rates are about 20\% more accurate than the estimate
of \citet{1994MNRAS.271..676L} for minor mergers and about three times
more accurate for minor mergers.  However, most of these studies have
focussed on the overall merger rates \citep{2008MNRAS.385.2025C}
rather than halo formation rates.

We discuss the Sasaki and the excursion set formalisms in \S 2.  We
describe our simulations in \S 3, discuss our results in \S 4 and
finally summarise our conclusions in \S 5.

\section{Rate of halo formation}
\label{sec:haloform}

The total change in number density of collapsed haloes at time $t$ with mass
between $M$ and $M+dM$ per unit time is denoted by $\dot{N}(M,t)dM$ and is due
to haloes gaining mass through accretion or mergers. 
Lower-mass haloes gain mass so that their mass is now between $M$ and $M+dM$,
and some of the haloes with mass originally between $M$ and $M+dM$ gain mass so
that their mass now becomes higher than this range.  
We call the former process halo formation and the latter as halo destruction,
even though the underlying physical process is the same in both cases; the
different labels of formation or destruction arise due to our perspective from
a particular range of mass.  
We denote the rate of halo formation by $\dot{N}_\mathrm{form}(M,t)dM$ and the
rate of halo destruction by $\dot{N}_\mathrm{dest}(M,t)dM$.  
We immediately have
\begin{equation}
 \dot{N}(M,t) = \dot{N}_\mathrm{form}(M,t)-\dot{N}_\mathrm{dest}(M,t).
 \label{eq:Ndot}
\end{equation}
Following \citet{1994PASJ...46..427S}, in general we can formulate each term
in the above expression as follows. 
The rate of halo destruction can be written as
\begin{eqnarray}
 \dot{N}_\mathrm{dest}(M,t) &=&
 \int\limits_{M}^{\infty}N(M,t)\tilde{Q}(M,M';t)dM' 
 \label{eq:Ndotdest1} \\
 &\equiv& \phi(M,t)N(M,t),
  \label{eq:Ndotdest2}
\end{eqnarray}
where, $\tilde{Q}(M,M';t)$ represents the probability of a halo of mass $M$
merging with another halo to form a new halo of mass $M'$ per unit time.  
The fraction of haloes that are destroyed per unit time is denoted by
$\phi(M,t)$.
This quantity is also referred to as the efficiency of halo destruction.
The rate of halo formation can be written as
\begin{equation}
 \dot{N}_\mathrm{form}(M,t) = \int\limits_{0}^{M}N(M',t)Q(M',M;t)dM'
  \label{eq:Ndotform1}
\end{equation}
where $Q(M',M;t)$ represents the probability of a halo of mass $M'$ evolving
into another halo of mass $M$ per unit time. 
We can now write, from equation (\ref{eq:Ndot}) and from our definitions in
equations (\ref{eq:Ndotdest2}) and (\ref{eq:Ndotform1}), 
\begin{equation}
 \dot{N}_\mathrm{form}(M,t) = \dot{N}(M,t)+\phi(M,t)N(M,t).
  \label{eq:Ndotform2}
\end{equation}
This reduces the calculation of rate of halo formation to a computation of
$\phi(M,t)$.  

\citet{1994PASJ...46..427S} proposed a simple ansatz to compute
$\phi(M,t)$: if we assume that the efficiency of halo destruction has
no characteristic mass scale and we require that the destruction rate remains
finite at all masses then it can be shown that $\phi$ does not depend on mass.

\subsection{Sasaki prescription: Press-Schechter mass function} 

To understand the Press-Schechter formalism \citep{1974ApJ...187..425P,
  1991ApJ...379..440B}, which gives the co-moving number density of collapsed
haloes at a time $t$ with mass between $M$ and $M+dM$, let us consider a dark
matter inhomogeneity centered around some point in the Universe. 
The smoothed density contrast within a smoothing scale of radius $R$ around
this point is defined as $\delta(R)=[\rho(R)-\bar\rho]/\bar\rho$, where
$\rho(R)$ is the density of dark matter within $R$ and $\bar\rho$ is the mean
background density of the Universe.  
If this density contrast $\delta(R)$ is greater than the threshold density
contrast for collapse $\delta_c$ obtained from spherical collapse model
\citep{1972ApJ...176....1G}, the matter enclosed within the volume collapses
to form a bound object.  
In hierarchical models, density fluctuations are larger at small scales so
with decreasing $R$, $\delta(R)$ will eventually reach $\delta_c$.  
The problem then is to compute the probability that the first up-crossing of
the barrier at $\delta_c$ occurs on a scale $R$. 
This problem can be addressed by excursion set approach. 

The excursion set approach consists of the following principles: consider a
trajectory $\delta(R)$ as a function of the filtering radius $R$ at any given
point and then determine the largest $R$ at which $\delta(R)$ up crosses the
threshold $\delta_c(t)$ corresponding to the formation time $t$. 
The solution of the problem can be enormously simplified for Brownian
trajectories \citep{1943RvMP...15....1C}, that is for sharp $k$-space filtered
density fields, as in this case contribution of each wave mode is independent
of all others.  
In such a case we have to solve the Fokker-Planck equation for the probability
density $\Pi(\delta,S)d\delta$, where $S\equiv \sigma^2(R)$ and $\sigma(R)$ is
the standard deviation of fluctuations in the initial density field, 
smoothed at a scale $R$,  
\begin{equation}
 \frac{\partial\Pi(\delta,S)}{\partial S} =
 \frac{1}{2}\frac{\partial^2\Pi(\delta,S)}{\partial\delta^2} 
\end{equation}
The solution \citep{1998MNRAS.298.1097P, 2007IJMPD..16..763Z} can be obtained
using the absorbing boundary condition $\Pi(\delta_c(t),S)=0$ and the initial
condition $\Pi(\delta,S=0)=\delta_{D}(\delta)$, where $\delta_{D}(\delta)$ is
the Dirac delta function  
\begin{equation}
\begin{split}
 \Pi(\delta,S;\delta_c)d\delta&=
  \frac{1}{\sqrt{2\pi S}}\\&
   \times\left[\exp\left(-\frac{\delta^2}{2S}\right) -
   \exp\left(-\frac{(\delta-2\delta_c(t))^2}{2S}\right)\right]d\delta
\end{split}   
\end{equation}
 Now define
 $F(S,\delta_c(t))=\int_{-\infty}^{\delta_c(t)}d\delta\Pi(\delta,S;\delta_c(t))$
 as the survival probability of trajectories and obtain the differential
 probability for a first barrier crossing:
\begin{equation}
 f(S)=-\frac{\partial F(S,\delta_c(t))}{\partial
   S}=\frac{\delta_c(t)}{\sqrt{2\pi
     S^3}}\exp\left(-\frac{\delta_c(t)^2}{2S}\right) 
\end{equation}
From this, one can obtain the co-moving number density of collapsed haloes at 
time $t$ with mass between $M$ and $M+dM$ 
\begin{eqnarray}
N_\mathrm{PS}(M,t)dM &=&
\frac{\rho_\mathrm{nr}}{M}f(S)\left|\frac{dS}{dM}\right|dM \nonumber\\ 
 &=& \sqrt{\frac{2}{\pi}} \frac{\rho_\mathrm{nr}}{M}
 \left(\nu\right)^\frac{1}{2}\left|\frac{d\ln\sigma}{dM}\right|
 \exp\left[-\frac{\nu}{2}\right]dM.   
\label{eq:N_PS}
\end{eqnarray}
here $\rho_\mathrm{nr}$ is the average comoving density of non-relativistic
matter and $\nu \equiv [\delta_{c}(t)/\sigma(M))]^2 \equiv
[\delta_{c}/(D(t)\sigma(M))]^2$, where $\delta_{c}$ is the threshold density
contrast for collapse, $D(t)$ is the linear rate of growth for density
perturbations and $\sigma(M) (\equiv S^{1/2})$ is the standard deviation of
fluctuations in the initial density field, which is smoothed over a scale that
encloses mass $M$. 

In the following discussion, we will denote the mass function by $N(M,t)$ if
the statement is independent of the specific form of the mass function.   
We will use a subscript $\mathrm{PS}$ when the statements apply only to the
Press-Schechter form of the mass function.  

With Sasaki's ansatz, the destruction rate efficiency $\phi$ can be written in
this case as 
\begin{equation}
 \phi(t) = \frac{1}{D(t)}\frac{dD(t)}{dt}.
  \label{eq:phiPS}
\end{equation} 
With this, we can write down the rate of halo formation for the
Press-Schechter mass function from equation (\ref{eq:Ndotform2}) as:  
\begin{eqnarray}
{\dot{N}}_\mathrm{form}(M,t) &=&
\dot{N}_\mathrm{PS}(M,t)+\frac{1}{D(t)}
\frac{dD(t)}{dt}N_\mathrm{PS}(M,t)\nonumber\\&=&\frac{1}{D(t)}\frac{dD(t)}{dt}  
N_\mathrm{PS}(M,t)
\frac{\delta_{c}^{2}}{\sigma^{2}(M)D^{2}(t)}. \label{eq:Ndotform3}  
\end{eqnarray}
Note that for haloes with large mass, that is in the limit
${\delta_{c}} \gg {\sigma(M)D(t)}$, $\dot{N}_\mathrm{form}$ approaches
$\dot{N}_\mathrm{PS}$.  
In other words, the total change in the number of haloes is determined by
formation of new haloes.  
For haloes with low mass, where $\sigma$ is much larger than unity, although
$\dot{N}_\mathrm{form}$ remains positive, the total change is dominated by
destruction and $\dot{N}_\mathrm{PS}$ becomes negative.    

We can also define two related, useful quantities now.
Firstly, the probability $p(t_{1},t_{2})$ that a halo which exists at $t_{1}$
continues to exist at $t_{2}$ without merging is given by   
\begin{equation}
 p(t_{1},t_{2}) =
 \exp\left[-\int\limits_{t_{1}}^{t_{2}}\phi(t')dt'\right]
 = \frac{D(t_{1})}{D(t_{2})}\qquad\mbox{ (where $t_{2}>t_{1}$) }
  \label{eq:survprob}
\end{equation}
This is usually known as the survival probability of haloes, and is
independent of halo mass in the Sasaki prescription. 
In this picture, the distribution of epochs $t_\mathrm{f}$ of formation of
haloes with mass $M$ at time $t$ is given by 
\begin{equation}
 F(M;t_\mathrm{f},t)dMdt_\mathrm{f} =
 \dot{N}_\mathrm{form}(M,t_\mathrm{f})p(t_\mathrm{f},t)dMdt_\mathrm{f}. 
  \label{eq:distribution}
\end{equation}

\subsection{Sasaki prescription: Sheth-Tormen mass function}
\label{sec:sasakist}

The Press-Schechter mass function does not provide a very good fit to
halo mass function obtained in N-body simulations.   
In particular, it under-predicts the number density of large mass haloes, and
over-predicts that of small mass haloes. 
Hence it is important to generalize the calculation of formation rates to
other models for mass function that are known to fit simulations better.  
The Sheth-Tormen form of mass function \citep{1999MNRAS.308..119S} is known to
fit simulations much better than the Press-Schechter form.\footnote{Even this
  form of halo mass function has poor accuracy in some cases, namely, for
  conditional mass functions with large mass ratios and for mass function in
  overdense regions \citep{2002MNRAS.329...61S}.  In applications involving
  these regimes it is perhaps advisable to use more accurate fitting functions
  to simulation data.  However, the Sheth-Tormen form still has the property
  of being considerably better than the Press-Schechter form while having a
  physical interpretation.  It is thus preferable in many semi-analytic models
  where the Press-Schechter form is used.} 
(For a comparison of both of these forms of halo mass function with
simulations, see Fig.~3 of \citealt{2001MNRAS.321..372J}.) 
The Sheth-Tormen mass function is given by
\begin{equation}
\begin{split}
 N_\mathrm{ST}(M,t)dM = A \sqrt{\frac{2}{\pi}}&
 \frac{\rho_\mathrm{nr}}{M}\left(a\nu\right)^{1/2}\left|
   \frac{d\ln\sigma}{dM}\right|\\& 
 \times\left[1+\left(a\nu\right)^{-p}\right]
 \exp\left[-\frac{a\nu}{2}\right]dM, 
\end{split}
\end{equation}
where the parameters $a$, $p$, and $A$ have best fit values of $a=0.707$,
$p=0.3$ and $A=0.322$ \citep{1999MNRAS.308..119S}, and the quantity $\nu$ is
as defined before.  
This form of mass function has the added advantage of being similar to the
mass function derived using a variable barrier motivated by ellipsoidal
collapse of overdense regions \citep{2001MNRAS.323....1S,
  2002MNRAS.329...61S}.  
Note that if we choose $A=0.5$, $p=0$ and $a=1$ then we recover the
Press-Schechter mass function derived using spherical collapse. 
Recently, it has been shown that the best fit values of these parameters
depend on the slope of the power spectrum \citep{2009arXiv0908.2702B}. 

We can now apply the Sasaki prescription to this form of mass function and
calculate the rates of halo formation and destruction
\citep{2007MNRAS.376..709R}. 
We get for the destruction rate efficiency
\begin{equation}
\phi(t) = \frac{1}{D}\frac{dD}{dt}\left[1-2p\right].\label{eq:phiST}
\end{equation}
Note that the destruction rate efficiency is independent of mass.
The rate of halo formation is then given by
\begin{equation}
 \dot{N}_\mathrm{form}^\mathrm{ST}(M,t) = -\frac{1}{D}\frac{dD}{dt}
 \left[\frac{2p}{1+\left(a\nu\right)^{-p}}-a\nu\right]
 N_\mathrm{ST}(M,t). \label{eq:NdotformST} 
\end{equation} 
Note that in this case, because of the extra term, the halo formation rate can
be negative for some values of halo mass. 
Since negative values of rate of halo formation are unphysical, this indicates
that the generalization of Sasaki approximation to the Sheth-Tormen mass
function is incorrect.   
The same problem is encountered if we use other models of the halo mass
function \citep{2009NewA...14..591S}.

However, since the basic framework outlined in the beginning of this section
is clearly correct, there should not be any problems in generalizing it to
other mass functions.  
It is therefore likely that the simplifying assumptions of the Sasaki method
that led to the estimate of the halo destruction rate efficiency of equation
(\ref{eq:phiST}) are responsible for negative halo formation rate.   

\subsection{Excursion set approach to halo formation rates: Press-Schechter mass function}

To check this assertion we perform an explicit calculation of the rate of halo
formation using the excursion set formalism. 
Recall that from equations (\ref{eq:Ndotdest1}) and (\ref{eq:Ndotdest2}), we
can write for the halo destruction rate efficiency as 
\begin{equation}
\phi(M_{1},t) =
\int\limits_{M_{1}}^{\infty}\tilde{Q}(M_{1},M_{2};t)dM_{2}, 
\label{eq:phigeneral1}
\end{equation}
where $\tilde{Q}(M_{1},M_{2};t)$ represents the probability that an object of
mass $M_1$ grows into an object of mass $M_2$ per unit time through merger or
accretion at time $t$.  
This quantity is also known as the transition rate.

In the excursion set formalism, the conditional probability for a halo of mass
$M_{1}$ present at time $t_{1}$ to merge with another halo to form a larger
halo of mass between $M_{2}$ and $M_{2}+dM_{2}$ at time $t_{2}>t_{1}$
\citep{1993MNRAS.262..627L, 1994MNRAS.271..676L} can be written for the
extended Press-Schechter mass function as 
\begin{equation}
\begin{split}
 f(M_{2},\delta_{2}&|M_{1},\delta_{1})dM_{2} =  \sqrt{\frac{2}{\pi}}
 \frac{\delta_{2}(\delta_{1}-\delta_{2})}{\delta_{1}}\sigma_{2}^{2}
 \left[\frac{\sigma_{1}^{2}}{
     \sigma_{2}^{2}(\sigma_{1}^{2}-\sigma_{2}^{2})}\right]^{\frac{3}{2}}\\
 &\times\exp\left[  
   -\frac{(\delta_{2}\sigma_{1}^{2}-\delta_{1} \sigma_{2}^{2})^{2}}{2
     \sigma_{1}^{2}\sigma_{2}^{2}(\sigma_{1}^{2}-\sigma_{2}^{2})}  
 \right]\left|\frac{d\sigma_{2}}{dM_{2}}\right|dM_{2}.\label{eq:condprob} 
\end{split}
\end{equation}
Here, $\sigma_1$ and $\sigma_2$ are values of the standard deviation of the
density perturbations when smoothed over scales that contain masses $M_1$ and
$M_2$ respectively, and $\delta_1$ and $\delta_2$ are the values of the
threshold density contrast for spherical collapse at time $t_1$ and $t_2$
respectively. 
Taking the limit $t_{2}$ tends to $t_{1}$, i.~e. $\delta_{2}$ tends to
$\delta_{1}$, we can determine the mean transition rate at time $t=t_1$: 
\begin{equation}
\begin{split}
\tilde{Q}(M_{1},M_{2};t)&dM_2 =
\sqrt{\frac{2}{\pi}}\sigma_{2}^{2} \left[
  \frac{\sigma_{1}^{2}}{\sigma_{2}^{2}(\sigma_{1}^{2}-\sigma_{2}^{2})}
\right]^{\frac{3}{2}}\left|\frac{d\delta}{dt}\right|\\ 
& \times\exp\left[ -\frac{\delta^{2}(\sigma_{1}^{2}-\sigma_{2}^{2})}{
    2\sigma_{1}^{2}\sigma_{2}^{2}}\right]\left| \frac{d\sigma_{2}}{dM_{2}}
\right|dM_{2}. 
\label{eq:transrate}
\end{split}
\end{equation}
This represents the probability that a halo of mass $M_{1}$ will accrete or
merge to form another halo of mass $M_{2}$ at time $t$. 
We can use this with equation (\ref{eq:phigeneral1}) to explicitly compute the
destruction rate, and hence the halo formation rate. 

However, in the excursion set method, an arbitrarily small change in the halo
mass is treated as creation of a new halo. 
As a result, the integral in equation (\ref{eq:phigeneral1}) diverges unless
we specify a ``tolerance'' parameter. 
We assume that a halo is assumed to have {\it survived}\/ unless its mass
increases such that $M_{1} \rightarrow M_2 \geq M_{1}(1+\epsilon)$ due to
either accretion or merging, where $\epsilon$ is a small number.   
This assumption allows us to introduce a lower cutoff in the integral in
equation (\ref{eq:phigeneral1}) and the lower limit changes to
$M_{1}(1+\epsilon)$, leading to a convergent integral. 
This is also physically pertinent for our application as infinitesimal changes
do not lead to variations in dynamical structure of haloes, and hence we do not
expect any changes in galaxies hosted in haloes that do not undergo a major
merger.  
This is similar in spirit to the assumption made elsewhere in the literature
that a halo is assumed to survive until its mass increases by a factor two
\citep{1994MNRAS.271..676L, 1996MNRAS.280..638K}.    
Note that N-body simulations have a natural cutoff due to the discrete nature
of N-body particles.  
With the introduction of this new parameter, the modified formula for the halo
destruction rate efficiency is given by 
\begin{equation}
 \phi(M_{1},t) =
 \int\limits_{M_{1}(1+\epsilon)}^{\infty}\tilde{Q}(M_{1},M_{2};t)dM_{2} 
  \label{eq:phigeneral2}
\end{equation}
This can then be used to calculate the rate of halo formation using equation
(\ref{eq:Ndotform2}). 

\begin{figure}
\centering
\includegraphics[width=2.8in, angle=0.0]{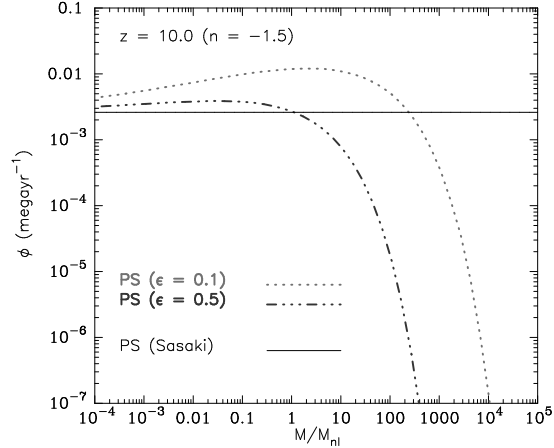}
  \caption{Destruction rate $\phi(M,t)$ at $z=10$ for the Press-Schechter mass
    function for a power law model with index $-1.5$.   Curves have been
    plotted for $\epsilon=0.1$ and $\epsilon=0.5$.} 
\label{fig:phi_ratio_n=-1.5_PS}
\end{figure}

Fig. \ref{fig:phi_ratio_n=-1.5_PS} shows the destruction rate efficiency
$\phi(M,t)$ computed in this manner for the Press-Schechter mass function for
an Einstein-de Sitter cosmology with power law spectrum of density
perturbations with index $-1.5$.    
Curves have been plotted for $\epsilon=0.1$ and $\epsilon=0.5$. 
We have also shown the Sasaki approximation in the same panel. 
The excursion set result has three features: 
\begin{enumerate}
\item
At small $M$, the excursion set value approaches the destruction rate
computed using the Sasaki approximation. 
\item
The destruction rate has a peak, more pronounced for smaller $\epsilon$, near
the scale of non-linearity. 
\item
At larger scales the destruction rate falls rapidly; this is the region where
deviations from the Sasaki result are the largest.  
Thus the halo destruction rate efficiency vanishes at large masses.
\end{enumerate}
A similar trend is seen for other power spectra. 
We postpone a detailed discussion of these issues to the end of this section.

\subsection{Excursion set approach to halo formation rates: Sheth-Tormen mass function}

As discussed in Subsection \ref{sec:sasakist}, the Sheth-Tormen mass function
is known to be a much better fit to N-body simulations than the
Press-Schechter mass function.  
Several other forms of halo mass function have also been fitted to results of
high resolution N-body simulations \citep{2001MNRAS.321..372J,
  2003MNRAS.346..565R, 2006ApJ...646..881W}.  
In this paper we focus on the Sheth-Tormen mass function.
Recall that the Sasaki prescription gave unphysical results when applied to
this form of the mass function. 
Therefore, we now derive the halo destruction rate efficiency, and the halo
formation rates for the Sheth-Tormen mass function. 
This requires obtaining analogs of equations (\ref{eq:condprob}) and
(\ref{eq:transrate}). 

\citet{2001MNRAS.323....1S} showed that once the barrier shape is known, all
the predictions of the excursion set approach, like the conditional mass
function, associated with that barrier can be computed easily\footnote{These
  can be calculated for non-Gaussian initial conditions, see, e.g.,
  \citet{2011arXiv1102.0046D}}.   
Further, they showed that the barrier shape for ellipsoidal collapse is 
\begin{equation}
 B(\sigma,t)\equiv\delta_\mathrm{ec}(\sigma,t) =
 \sqrt{a}\delta_{c}(t)\left[1+\beta(a\nu)^{-\gamma}\right],\label{eq:barrierST} 
\end{equation}
where $a = 0.75$, $\beta = 0.485$, $\gamma = 0.615$, and, $\delta_{c}(t)$ is
the threshold value of overdensity required for spherical collapse (also see
\citealt{2002MNRAS.329...61S}). 
They also found that, for various barrier shapes $B(S)$, the first-crossing
distribution of the excursion set theory is well approximated by   
\begin{equation}
f(S)dS = \frac{|T(S)|}{\sqrt{2\pi S^{3/2}}}
\exp\left[-\frac{B(S)^{2}}{2S}\right]dS,\label{eq:fSST} 
\end{equation}
where $T(S)$ denotes the sum of the first few terms in the Taylor series
expansion of $B(S)$ 
\begin{equation}
T(S) = \sum\limits_{n=0}^{\infty}
\frac{(-S)^{n}}{n!}\frac{\partial^{n}B(S)}{\partial S^{n}}.\label{eq:TS} 
\end{equation}
(Here, for conformity with the literature, we use the symbol
$S\equiv\sigma^2$.)  
This expression gives the exact answer in the case of constant and linear
barriers.  
For the ellipsoidal barrier, we can get convergence of the numerical
result if we retain terms in the Taylor expansion up to $n = 5$. 

For Press-Schechter mass function, the conditional mass function
$f(S_{1},\delta_{1}|S_{2},\delta_{2})$ can be obtained from the first
crossing $f(S)$ by just changing the variables
$\delta\rightarrow\delta_{1}-\delta_{2}$ and $S\rightarrow
S_{1}-S_{2}$.  This can be done because, despite the shift in the
origin, the second barrier is still one of constant height.  This is
no longer true for Ellipsoidal collapse and hence we cannot simply
rescale the function of equation to get the conditional mass function.
Instead, this can be done by making the replacements $B(S) \rightarrow
B(S_{1})-B(S_{2})$ and $S\rightarrow S_{1}-S_{2}$ in equation (\ref{eq:fSST}).  
\begin{equation}
\begin{split}
 f(S_{1}|S_{2})dS_{1} =
 &\frac{|T(S_{1}|S_{2})|}{\sqrt{2\pi(S_{1}-S_{2})^{3/2}}} \\
 &\times\exp\left[-\frac{
     \left(B(S_{1})-B(S_{2})\right)^{2}}{2(S_{1}-S_{2})}\right]
 dS_{1}, \label{eq:fS1S2} 
\end{split}
\end{equation}
where we now have
\begin{equation}
 T(S_{1}|S_{2}) = \sum\limits_{n=0}^{5} \frac{(-(S_{1}-S_{2}))^{n}}{n!}
 \frac{\partial^{n}\left(B(S_{1})-B(S_{2})\right)}{\partial
   S^{n}_{1}}.  \label{eq:TS1S2} 
\end{equation}
Using Bayes' theorem, we now have  
\begin{equation}
\begin{split}
 f&\left(S_{2}|S_{1}\right)dS_{2} = \frac{|T(S_{1}|S_{2})|
   |T(S_{2})|}{|T(S_{1})|}
 \frac{1}{\sqrt{2\pi}}\left[\frac{S_{1}}{S_{2}\left(S_{1}-S_{2}\right)}\right]
 \\ &\times\exp\left[
   -\frac{\left[B(S_{1})-B(S_{2})\right]^{2}}{2\left(S_{1}-S_{2}\right)}-
   \frac{B^{2}(S_{2})}{2S_{2}}+\frac{B^{2}(S_{1})}{2S_{1}}\right]dS_{2}. 
\label{eq:fS2S1}  
\end{split}
\end{equation}
A change of variables from $S$ to $M$ now gives us an analog of equation
(\ref{eq:condprob}) for the Sheth-Tormen mass function.  In other words, we
get the conditional probability $f_\mathrm{ST}(M_{2}|M_{1})d\ln M_{2}$ that a
halo of mass $M_{1}$ present at time $t_{1}$ will merge to form a halo of mass
between $M_{2}$ and $M_{2}+dM_{2}$ at time $t_{2}>t_{1}$. 
Further, taking the limit as $t_{2}$ tends to $t_{1} (= t)$, we obtain
$\tilde{Q}(M_{1},M_{2};t)$.   As before, we can then use it to calculate the
halo destruction rate efficiency $\phi(M,t)$ and the rate of halo formation
$\dot{N}_\mathrm{form}^\mathrm{ST}(M_{1},z)$ using equations
(\ref{eq:Ndotform2}) and (\ref{eq:phigeneral2}).   
We perform this part of the calculation numerically.
It is also possible to use this formalism to calculate formation rates for the
square-root barrier \citep{2009MNRAS.397..299M, 2008MNRAS.391.1729M,
  2007MNRAS.376..977G}, which is a good approximation for the ellipsoidal
collapse model.  
We do not attempt this calculation here as it is beyond the scope of this
paper.  

\begin{figure}
  \centering
  \includegraphics[width=2.8in, angle=0.0]{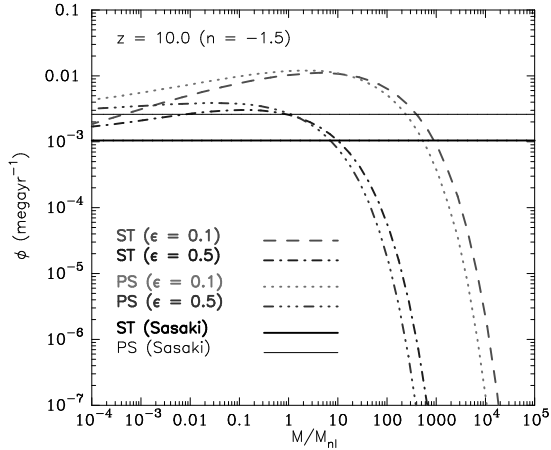}
  \caption{Same as Fig.~\ref{fig:phi_ratio_n=-1.5_PS} but for the
    Sheth-Tormen (as well as Press-Schechter) mass function.}
  \label{fig:phi_ratio_n=-1.5_ST}
\end{figure}

Fig. \ref{fig:phi_ratio_n=-1.5_ST} is the analog of Fig.
\ref{fig:phi_ratio_n=-1.5_PS} for the Sheth-Tormen mass function.   
It shows the destruction rate per halo $\phi(M,t)$ computed using the
excursion set method for an Einstein-de Sitter cosmology with power law
spectrum of density perturbations with index $-1.5$ at $z=10.0$.     
Curves have been plotted for $\epsilon=0.1$ and $\epsilon=0.5$. 
We have also shown the Sasaki approximation for ST mass function in the same
panel for comparison. 
This result for the Sheth-Tormen mass function has the same features as the
result for the Press-Schechter mass function.   
We also see that the destruction rate efficiency is far from constant at small
$M/M_{nl}$.   
Thus the central assumption of the Sasaki prescription is invalid in the case
of Sheth-Tormen mass function as well. 

\begin{table}
\caption{For power law: here $n$ is the index of the power spectrum,
  $N_\textrm{box}$ is the size of the simulation box , $N_\textrm{part}$ represents
  the number of particles, $\rnl^i$ is the scale of non-linearity at the
  earliest epoch, $\rnl^f$ is the actual scale of non-linearity for the last
  epoch, $\rnl^\textrm{max}$ represents the maximum scale of non-linearity and
  $z_i$ is the starting redshift of the simulations for every model.}
\begin{center}
\begin{tabular}{ccccccc}
\hline
$n$ & $N_\mathrm{box}$ & $N_\mathrm{part}$ & $\rnl^i$ & $\rnl^f$ &
$\rnl^\mathrm{max}$ & $z_i$\\ 
\hline
$-1.5$ & $400^3$ & $400^3$ & $2.5$ & $12.0$ & $10.0$ & $103.38$\\
$-0.5$ & $256^3$ & $256^3$ & $2.5$ & $12.0$ & $18.2$ & $291.53$\\
\hline
\label{table_nbody_runs_powlaw}
\end{tabular}
\end{center}
\end{table}

\begin{table}
\caption{For LCDM: columns 1 and 2 list the size of the box (in Mpc$/h$) and the
  number of particles used in the simulations. Columns 3 and 4 give the mass
  (in M$_\odot/h$) and force resolution (in kpc$/h$; not to be confused with
  the $\epsilon$ used in the text) of the simulations, while columns 5 and 6
  tell us the redshift at  which  the simulations were terminated and the
  redshift for which the analyses were done.}
\begin{center}
\begin{tabular}{cccccc}
\hline
$L_\mathrm{box}$ & $N_\mathrm{part}$ & $m_\mathrm{part}$ & $\epsilon$ &
$z_\mathrm{f}$ & $z_\mathrm{out}$\\ 
\hline
$23.04$ & $512^3$ & $6.7 \times 10^6$ & 1.35 & 5.0 & 5.04 \\
$51.20$ & $512^3$ & $7 \times 10^7$ & 3.00 & 3.0 & 3.34 \\
$76.80$ & $512^3$ & $2.3 \times 10^8$ & 4.50 & 1.0 & 1.33 \\
\hline
\label{table_nbody_runs_lcdm}
\end{tabular}
\end{center}
\end{table}

\section{N-body simulations}
\label{sec:simulations}

From the excursion set calculation described in the previous section,
we thus find that the halo destruction rate efficiency is not
independent of mass as is assumed in the Sasaki prescription.
Clearly, this is the reason why Sasaki prescription yields unphysical
values for the rate of halo formation.  In this section and the next,
we now compare the results of our excursion set calculation with
results of N-body simulations.

We used the TreePM code \citep{2009RAA.....9..861K} for these simulations.
The TreePM \citep{2002JApA...23..185B, 2003NewA....8..665B} is a
hybrid N-body method which improves the accuracy and performance of
the Barnes-Hut (BH) Tree method \citep{1986Natur.324..446B} by
combining it with the PM method \citep{1983ApJ...270..390M,
  1983MNRAS.204..891K, 1985A&A...144..413B, 1985ApJ...299....1B,
  1988csup.book.....H, 1997Prama..49..161B, 2005NewA...10..393M}.
The TreePM method explicitly breaks the potential into a short-range
and a long-range component at a scale $r_s$: the PM method is used to
calculate the long-range force and the short-range force is computed using
the BH Tree method.
Use of the BH Tree for short-range force calculation enhances the force
resolution as compared to the PM method.

The mean interparticle separation between particles in the simulations
used here is $l_\mathrm{mean} = 1.0 $ in units of the grid-size used
for the PM part of the force calculation.
In our notation this is also cube root of the ratio of simulation volume
$N_\mathrm{box}^3$ to the total number of particles $N_\mathrm{part}$.

Power law models do not have any intrinsic scale apart from the scale of
non-linearity introduced by gravity. 
We can therefore identify an epoch in terms of the scale of
non-linearity $\rnl$.
This is defined as the scale for which the linearly extrapolated 
value of the mass variance at a given epoch $\sigma_L(a,\rnl)$ is
unity.
All power law simulations are normalized such that $\sigma^2(a=1.0,\rnl=8.0) =
1.0$.  
The softening length in grid units is $0.03 $ in all runs.

The $\Lambda$CDM simulations were run with the set of cosmological parameters
favored by {\it Wilkinson Microwave Anisotropy Probe} 5-yr data (WMAP;
\citealt{2009ApJS..180..330K}) as the best fit for the $\Lambda$CDM class of
models: $\Omega_{nr} = 0.2565, \Omega_{\Lambda} = 0.7435, n_s = 0.963,
\sigma_8 = 0.796, h = 0.719$ and $\Omega_bh^2 = 0.02273$. 
The simulations were done with $512^3$ particles in a comoving cube of three
different values of the physical volume as given in Table
\ref{table_nbody_runs_lcdm}.  

Simulations introduce an inner and an outer scale in the problem and in most
cases we work with simulation results where $L_\mathrm{box} \gg \rnl \geq
L_\mathrm{grid}$, where $L_\mathrm{grid}$, the size of a grid cell is the
inner scale in the problem. 
$L_\mathrm{box}$ is the size of the simulation and represents the outer scale.
In Table (\ref{table_nbody_runs_powlaw}) we list the power law models
simulated for the present study.
We list the index of the power spectrum $n$ (column 1), size of the
simulation box $N_\textrm{box}$ (column 2), number of particles
$N_\textrm{part}$ (column 3), the scale of non-linearity at the earliest
epoch used in this study (column 4), and, the maximum scale of
non-linearity, $\rnl^\textrm{max}$ (column 6) given our tolerance level
of $3\%$ error in the mass variance at this scale.  
For some models with very negative indices we have run the simulations beyond
this epoch.  
This can be seen in column 5 where we list the actual scale of non-linearity
for the last epoch. 
The counts of haloes in low mass  bins are relatively unaffected  by finite
box considerations.  
We therefore limit errors in the mass function by running the simulation 
up to $\rnl^\mathrm{max}$ .
Column 7 lists the starting redshift of the simulations for every model. 
Similarly, in Table (\ref{table_nbody_runs_lcdm}), we mention the details of 
the LCDM simulations used in this work. We list the size of the simulation box 
$L_\mathrm{box}$ in Mpc$/h$ (column 1), number of particles used in the 
simulations $N_\mathrm{part}$ (column 2), mass of the particles
$m_\mathrm{part}$ in M$_\odot/h$ (column 3), force resolution $\epsilon$ (not
to be confused with 
the $\epsilon$ used in the text) of the simulations in kpc$/h$ (column 4), 
the redshift $z_\mathrm{f}$ at which the simulations were terminated (column 5) 
and the redshift $z_\mathrm{out}$ for which the analyses were done (column 6). 

In order to follow the merger history of dark matter haloes in each of these
simulations, we store the particle position and velocities at different
redshifts.  
A friend-of-friend group finding algorithm is used to locate the virialised
haloes in each of these slices.  
We adopt a linking length that is $0.2$ times the mean inter-particle
separation, corresponding to the density of virialised haloes.  
Only groups containing at least $20$ particles are included in our halo
catalogs. 
A merger tree is then constructed out of the halo catalogs by tracking the
evolution of each particle through various slices. 
This lets us identify a halo as it evolves with time through mergers with
other haloes. 
We then describe the formation and destruction of haloes in terms of change in
number of particles between consecutive snapshots of the simulation. 
When a halo of mass $M$ at redshift $z$ turns into a halo of mass $M'$ at
$z'(<z)$, then we say that a halo of mass $M$ was destroyed at redshift $z$
and a halo of mass $M'$ has formed at $z'$ if $M' \ge M(1+\epsilon)$.  We
identify the resolution parameter $\epsilon$ with that used in our excursion
set calculation and experiment with different values as described in the
next section. 

We find that a tolerence parameter $\epsilon$, similar to the one
defined before, also occurs while analysing the results of N-body
simulations.  We identify these two quantities.  As we will see in the
next section, the formation rate in our model has a dependence on
$\epsilon$, which reproduces the dependence of the results of N-body
simulations on this quantity.  Thus, the presence of $\epsilon$ in our
analytical model is crucial in comparing our results with the N-body
results.

\begin{figure}
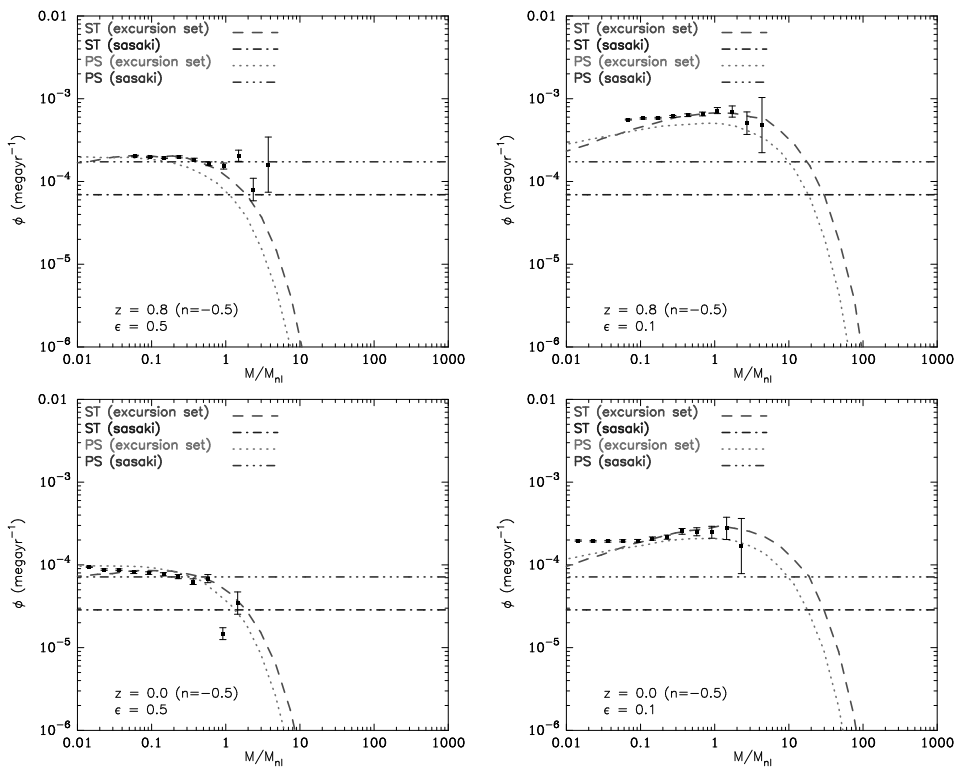

  \centering
\begin{tabular}{cc}
  \includegraphics[scale=.4]{./phi_n=-0.5_z=0.8_eps=0.5.ps} &
  \includegraphics[scale=.4]{./phi_n=-0.5_z=0.8_eps=0.1.ps} \\
  \includegraphics[scale=.4]{./phi_n=-0.5_z=0.0_eps=0.5.ps} &
  \includegraphics[scale=.4]{./phi_n=-0.5_z=0.0_eps=0.1.ps} \\
\end{tabular}
  \caption{Comparison of the destruction rate efficiencies computed
    using our method and Sasaki formalism for both ST and PS mass
    function at $r_{nl}=5$ grid lengths (top row) and $r_{nl}=8$ grid
    lengths (second row).  All curves are plotted for power-law model
    with index $n=-0.5$. Curves for $\epsilon=0.5$ are shown in the
    left panel and $\epsilon=0.1$ in the right panel.  Points with
    error bars represent the corresponding results obtained from
    N-body simulations.}
  \label{fig:n=-0.5}
\end{figure}

\begin{figure}
  \centering
\begin{tabular}{cc}
  \includegraphics[scale=.4]{./phi_n=-1.5_z=0.7_eps=0.5.ps} &
  \includegraphics[scale=.4]{./phi_n=-1.5_z=0.7_eps=0.1.ps} \\
  \includegraphics[scale=.4]{./phi_n=-1.5_z=0.0_eps=0.5.ps} &
  \includegraphics[scale=.4]{./phi_n=-1.5_z=0.0_eps=0.1.ps} \\
\end{tabular}
  \caption{Same as Fig.~\ref{fig:n=-0.5} but now for 
   $n=-1.5$.  The two epochs correspond to $r_{nl}=4$ and $r_{nl}=8$ grid
   lengths respectively.}
  \label{fig:n=-1.5}
\end{figure}

\begin{figure}
\centering 
\begin{tabular}{cc}
  \includegraphics[scale=.4]{./phi_lcdm_z=10.2_eps=0.5.ps} &
  \includegraphics[scale=.4]{./phi_lcdm_z=2.0_eps=0.5.ps} \\
  \includegraphics[scale=.4]{./phi_lcdm_z=10.2_eps=0.1.ps} &
  \includegraphics[scale=.4]{./phi_lcdm_z=2.0_eps=0.1.ps} \\
\end{tabular}
  \caption{Destruction rates for $\Lambda$CDM model for both PS and ST
    mass functions using different thresholds ($\epsilon=0.5$ for
    first; $\epsilon=0.1$ for second) and different redshifts
    ($z=10.2$ for left panel, $z=2.0$ for right panel). Again, points
    with error bars represent the corresponding results obtained from
    N-body simulations.}
  \label{fig:lcdm}
\end{figure}

\section{Results and discussion}

In this section we present the results of a comparison of our calculations
presented in Section \ref{sec:haloform} with N-body simulations.   
We present comparison of the destruction rate efficiency and the rate of halo
formation and then discuss our results at the end of this section. 
We also consider two related quantities, the halo survival probability and the
distribution of halo formation times, that were defined in Section
\ref{sec:haloform}. 

\begin{figure}
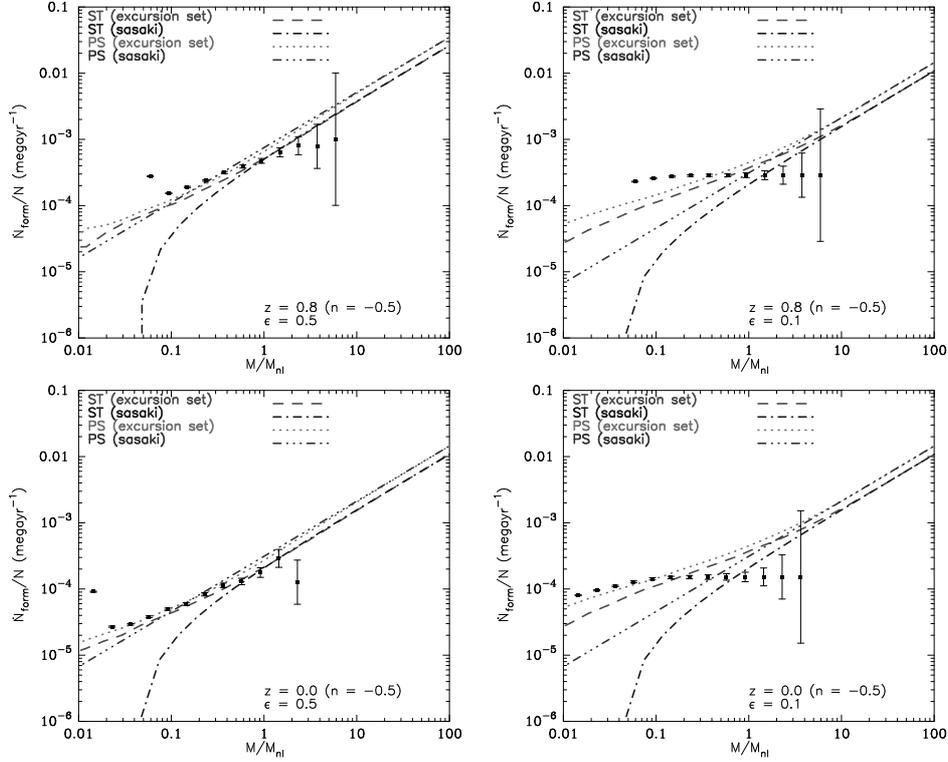

  \centering
\begin{tabular}{cc}
  \includegraphics[scale=.4]{./formation_rate_n=-0.5_z=0.8_eps=0.5.ps} & 
  \includegraphics[scale=.4]{./formation_rate_n=-0.5_z=0.8_eps=0.1.ps} \\
  \includegraphics[scale=.4]{./formation_rate_n=-0.5_z=0.0_eps=0.5.ps} &
  \includegraphics[scale=.4]{./formation_rate_n=-0.5_z=0.0_eps=0.1.ps}
\end{tabular}
  \caption{Comparison of the formation rates computed using our method
    and Sasaki formalism for both ST and PS mass functions for
    $r_{nl}=5$ (top row) and $r_{nl}=8$ (bottom row). All curves are
    plotted for power-law model with index $n=-0.5$. Curves for
    $\epsilon=0.5$ are shown in the left panel and $\epsilon=0.1$ in
    the right panel. Points with error bars represent the
    corresponding results obtained from N-body simulations.}
  \label{fig:n=-0.5_formation}
\end{figure}

\begin{figure}
  \centering
\begin{tabular}{cc}
  \includegraphics[scale=.4]{./formation_rate_n=-1.5_z=0.7_eps=0.5.ps} &
  \includegraphics[scale=.4]{./formation_rate_n=-1.5_z=0.7_eps=0.1.ps} \\
  \includegraphics[scale=.4]{./formation_rate_n=-1.5_z=0.0_eps=0.5.ps} &
  \includegraphics[scale=.4]{./formation_rate_n=-1.5_z=0.0_eps=0.1.ps} \\
\end{tabular}
  \caption{Same as Fig.~\ref{fig:n=-0.5_formation} but now for 
   $n=-1.5$.  The two epochs correspond to $r_{nl}=4$ and $r_{nl}=8$ grid
   lengths respectively.}
  \label{fig:n=-1.5_formation}
\end{figure}

\subsection{Halo destruction rate efficiency}

Figs.~\ref{fig:n=-0.5} and \ref{fig:n=-1.5} show the halo destruction rate
efficiency $\phi(M,t)$ for Sheth-Tormen and Press-Schechter mass functions in
an Einstein-de Sitter universe with a power law power spectrum of density
fluctuations with indices $n=-0.5$ and $n=-1.5$ respectively.   
The top row of both figures shows the halo destruction rate efficiency at
$z=0.8$ and the second row shows the same at $z=0.0$.   
In each case, we compute the halo destruction rate efficiency using the Sasaki
method as well as our excursion set method.   
We then derive $\phi(M,t)$ from our N-body simulations for a comparison: see 
\citet{2009arXiv0908.2702B} for details of the simulations and best fit
parameters for the ST mass function. 
These results are superimposed on the plots.
For the excursion set calculation and for the comparison with simulations, we
use $\epsilon=0.5$ (left column) and $\epsilon=0.1$ (right column).   
For the two power spectra, the two redshifts that we consider correspond to
$r_\mathrm{nl}=5$ and $r_\mathrm{nl}=8$ grid lengths, and $r_\mathrm{nl}=4$
and $r_\mathrm{nl}=8$ grid lengths respectively. 

As we saw in Figs.~\ref{fig:phi_ratio_n=-1.5_PS} and
\ref{fig:phi_ratio_n=-1.5_ST}, we find that Sasaki's assumption is not valid
for ST or PS mass functions, that is \ $\phi(M,t)$ depends on the halo mass. 
We also see that the value of $\phi(M,t)$ derived from simulations matches
well with that calculated by our method. 
On the other hand, the predictions of Sasaki's approximation do not match the
simulations. 
This difference is more pronounced for the smaller value of $\epsilon$. 
Note that the points from N-body simulations have large error-bars at higher
mass as the number of haloes decreases at these scales. 
The most notable feature of the destruction rate efficiency in the excursion
set picture is that it cuts off very sharply for large masses.  
Another aspect is that for small $\epsilon$, there is a pronounced peak in
$\phi$ and it drops off towards smaller masses.  

We have also calculated the destruction rate efficiency for the
$\Lambda$CDM cosmological model for both Press-Schechter and
Sheth-Tormen mass functions and compared it with derived values from
simulations.  The results are shown in Fig. \ref{fig:lcdm} for two 
redshifts ($2.0$ and $10.2$) and two values of $\epsilon$ ($0.5$ and
$0.1$).  We can see that results calculated by our technique fit
numerical results better.

\begin{figure}
\centering 
\begin{tabular}{cc}
  \includegraphics[scale=.4]{./formation_rate_lcdm_z=10.2_eps=0.5.ps} &
  \includegraphics[scale=.4]{./formation_rate_lcdm_z=2.0_eps=0.5.ps} \\
  \includegraphics[scale=.4]{./formation_rate_lcdm_z=10.2_eps=0.1.ps} &
  \includegraphics[scale=.4]{./formation_rate_lcdm_z=2.0_eps=0.1.ps} \\
\end{tabular}
  \caption{Formation rates for $\Lambda$CDM model for both PS and ST
    mass functions using different thresholds ($\epsilon=0.5$ for
    first; $\epsilon=0.1$ for second row) and different redshifts
    ($z=10.2$ for left panel, $z=2.0$ for right panel). As usual,
    points with error bars represent the corresponding results
    obtained from N-body simulations.}
  \label{fig:lcdm_formation}
\end{figure}

\begin{figure}
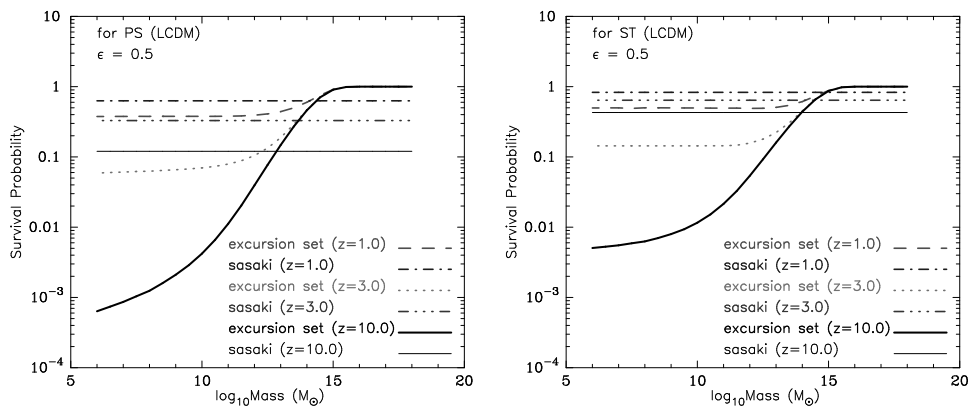

  \centering
\begin{tabular}{cc}
  \includegraphics[width=2.4in, angle=0.0]{./plot_survival_prob_PS_LCDM_all_eps0.5.ps} &
  \includegraphics[width=2.4in, angle=0.0]{./plot_survival_prob_ST_LCDM_all_eps0.5.ps} 
\end{tabular}
  \caption{Comparison of the survival probabilities computed using our method
    and Sasaki formalism for both PS (left panel) and ST (right panel) mass
    functions with different redshifts ($z=1$, $3$ and $10$) for
    $\epsilon=0.5$. Curves have been plotted for the $\Lambda$CDM model.  These
    curves show the probability that the halo survives from that redshift up
    to the present epoch.} 
  \label{fig:survival_prob}
\end{figure}

\begin{figure}
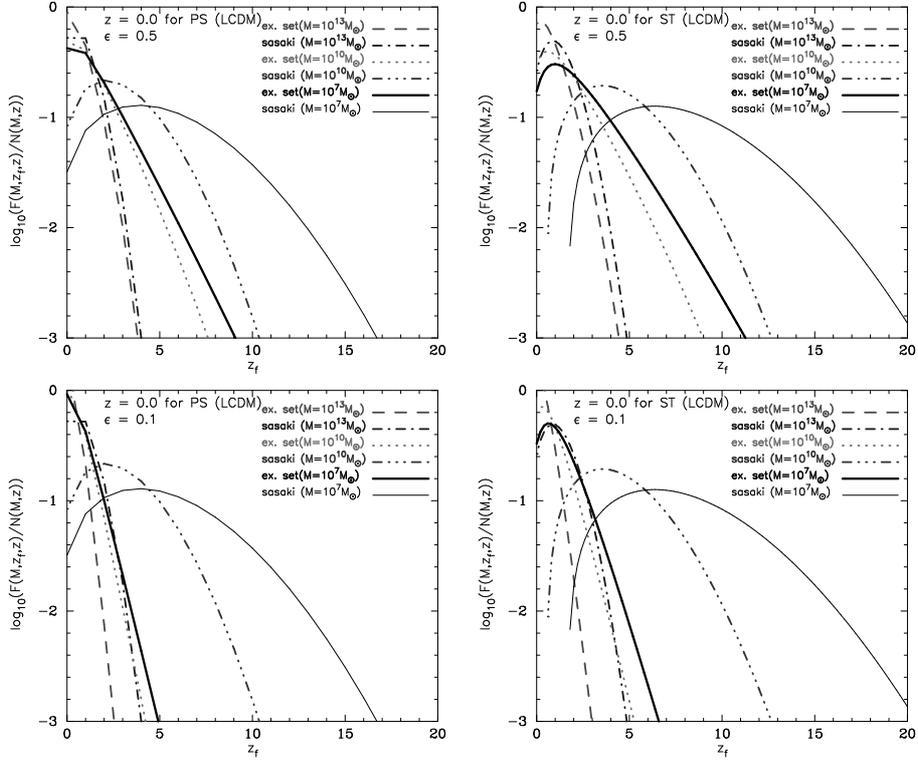

  \centering
\begin{tabular}{cc}
  \includegraphics[scale=.4]{./plot_form_dist_PS_LCDM_all_eps0.5.ps} &
  \includegraphics[scale=.4]{./plot_form_dist_ST_LCDM_all_eps0.5.ps}\\
  \includegraphics[scale=.4]{./plot_form_dist_PS_LCDM_all_eps0.1.ps} &
  \includegraphics[scale=.4]{./plot_form_dist_ST_LCDM_all_eps0.1.ps}\\
\end{tabular}
  \caption{Plots for formation epoch distribution of haloes.  Left column is for
  the PS and the right column is for the ST mass function. Curves have been
  plotted for the LCDM model.  The formation epoch distribution as computed
  using the Sasaki formalism and the excursion set approach described in this
  work is shown in the top panel.}
  \label{fig:formation_epoch_distribution}
\end{figure}

\begin{figure}
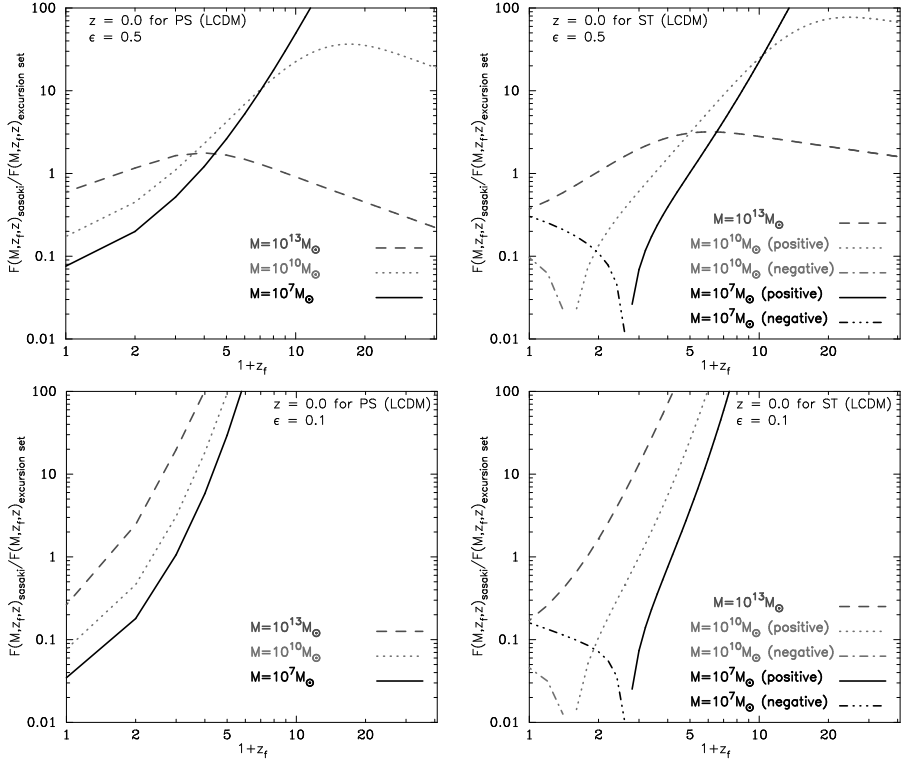

  \centering
\begin{tabular}{cc}
  \includegraphics[scale=.4]{./plot_form_dist_PS_LCDM_all_eps0.5_ratio.ps} &
  \includegraphics[scale=.4]{./plot_form_dist_ST_LCDM_all_eps0.5_ratio.ps}\\   
  \includegraphics[scale=.4]{./plot_form_dist_PS_LCDM_all_eps0.1_ratio.ps} &
  \includegraphics[scale=.4]{./plot_form_dist_ST_LCDM_all_eps0.1_ratio.ps}
\end{tabular}
  \caption{Ratio of the two different approaches used in Fig.
    \ref{fig:formation_epoch_distribution} to highlight less obvious
    differences.}
  \label{fig:formation_epoch_distribution_ratio}
\end{figure}

\begin{figure}
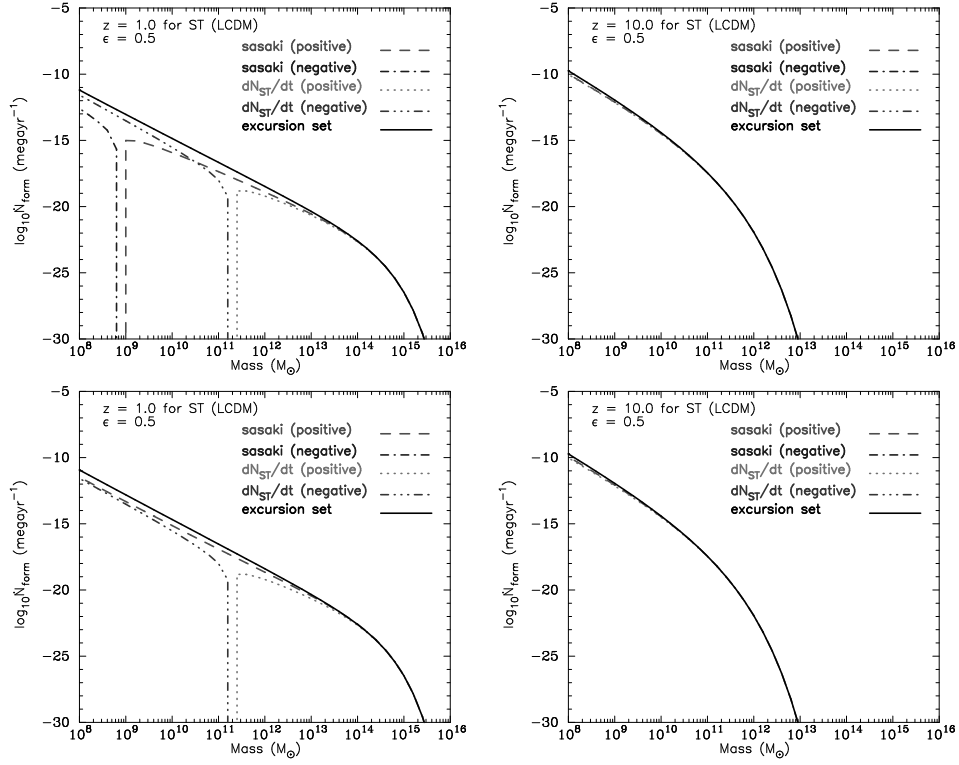

  \centering
\begin{tabular}{cc}
  \includegraphics[scale=.4]{./compare_ndotform_using_phi_sasaki_ST_z=1_eps=0.5.ps} &
  \includegraphics[scale=.4]{./compare_ndotform_using_phi_sasaki_ST_z=10_eps=0.5.ps} \\
  \includegraphics[scale=.4]{./compare_ndotform_using_phi_sasaki_PS_z=1_eps=0.5.ps} &
  \includegraphics[scale=.4]{./compare_ndotform_using_phi_sasaki_PS_z=10_eps=0.5.ps} \\
\end{tabular}
  \caption{Upper panels show formation rates for ST mass function.
    Lower panels show the same where we used $\phi$ computed from
    excursion set approach in the PS mass function and used that to
    compute the formation rate in the ST mass function.}
  \label{fig:lcdm_approx_eps0.5}
\end{figure}

\begin{figure}
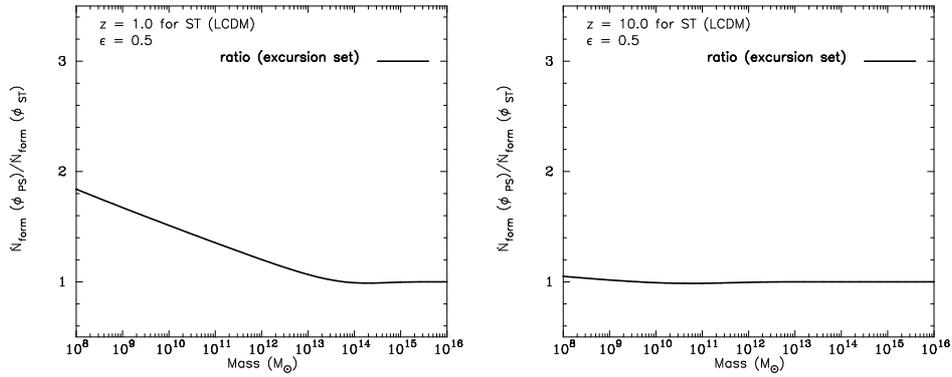

  \centering
\begin{tabular}{cc}
  \includegraphics[scale=.4]{./formation_rate_lcdm_test_z=1.0_eps=0.5.ps} &
  \includegraphics[scale=.4]{./formation_rate_lcdm_test_z=10.0_eps=0.5.ps} 
\end{tabular}
  \caption{Ratio of formation rates estimated in the two approaches
    shown in Fig.~\ref{fig:lcdm_approx_eps0.5}.}
  \label{fig:lcdm_approx_eps0.5_ratio}
\end{figure}

\begin{figure}
  \centering
\begin{tabular}{cc}
  \includegraphics[scale=.4]{./compare_ndotform_using_phi_sasaki_ST_z=1_eps=0.1.ps} &
  \includegraphics[scale=.4]{./compare_ndotform_using_phi_sasaki_ST_z=10_eps=0.1.ps} \\
  \includegraphics[scale=.4]{./compare_ndotform_using_phi_sasaki_PS_z=1_eps=0.1.ps} &
  \includegraphics[scale=.4]{./compare_ndotform_using_phi_sasaki_PS_z=10_eps=0.1.ps} \\
\end{tabular}
  \caption{Same as Fig.~\ref{fig:lcdm_approx_eps0.5} but for $\epsilon=0.1$}
  \label{fig:lcdm_approx_eps0.1}
\end{figure}

\begin{figure}
  \centering
\begin{tabular}{cc}
  \includegraphics[scale=.4]{./formation_rate_lcdm_test_z=1.0_eps=0.1.ps} &
  \includegraphics[scale=.4]{./formation_rate_lcdm_test_z=10.0_eps=0.1.ps} \\
\end{tabular}
  \caption{Same as Fig.~\ref{fig:lcdm_approx_eps0.5_ratio} but for $\epsilon=0.1$}
  \label{fig:lcdm_approx_eps0.1_ratio}
\end{figure}

\subsection{Halo formation rate} 

Having calculated the destruction rate efficiency, we can now
calculate the halo formation rate using the formalism described in
Section \ref{sec:haloform} and compare it with the derived halo
formation rates from our simulations.  The results are shown in
Figs.~\ref{fig:n=-0.5_formation} and \ref{fig:n=-1.5_formation} for
an Einstein-de Sitter Universe with a power law power spectrum of
density fluctuations with indices $n=-0.5$ and $n=-1.5$ respectively.
The third row of both figures shows the formation rate at redshift
$z=0.8$ and the fourth row shows the same at redshift $z=0.0$.  Note
the quantity plotted here is the ratio $\dot
N_\mathrm{form}(M,t)/N(M,t)$.  We have shown the results from the
Sasaki prescription and the excursion set calculations and have
superimposed formation rates derived from N-body simulations.  As
before, for the excursion set calculation and for the comparison with
simulations, we use $\epsilon=0.5$ (left column) and $\epsilon=0.1$
(right column).  For the two power spectra, the two redshifts that we
consider correspond to $r_\mathrm{nl}=5$ and $r_\mathrm{nl}=8$ grid
lengths, and $r_\mathrm{nl}=4$ and $r_\mathrm{nl}=8$ grid lengths
respectively. The corresponding results for the $\Lambda$CDM 
cosmological model are shown in Fig.~\ref{fig:lcdm_formation} for 
two redshifts (2.0 and 10.2) and two values of $\epsilon$ (0.5 and 0.1).

Again, we see that the excursion set results fit simulation data much
better as compared to the results from Sasaki prescription.  The
Sasaki method underestimates the formation rates by a large factor for
low mass haloes.  Results from the two methods tend to converge in the
large mass limit, although a systematic difference remains between the
Sheth-Tormen and Press-Schechter estimates, with the former always
being larger that the later.  The difference in the Sasaki estimate
and the excursion set estimate for the destruction rate efficiency and
the formation rate is as high as an order of magnitude at some scales
so the close proximity of simulation points to the excursion set
calculations is a clear vindication of our approach.  It is worth
noting that there is a clear deviation of simulation points from the
theoretical curves at small mass scales and this deviation is more
pronounced at small mass scales for $\epsilon=0.5$.  It may be that
some of the deviations arise due to a series representation of the
barrier shape, and the number of terms taken into account may not
suffice for the estimate.  We have found that truncation of the series
can affect results at small masses, though in most cases results
converge with the five terms that we have taken into account for the
range of masses considered here.

\subsection{Halo survival probability}

An important auxiliary quantity in the ongoing discussion is the halo
survival probability, defined in Section \ref{sec:haloform}.  From our
calculation of the halo destruction rate efficiency, we calculated the
survival probability of dark matter haloes using both the excursion set
formalism and the Sasaki prescription and compared results.  These
results are shown in Fig.~\ref{fig:survival_prob}, which shows the
survival probabilities in the $\Lambda$CDM cosmological model for the
Press-Schechter (left panel) and Sheth-Tormen (right panel) mass
functions using the two approaches at three different redshifts
($z=1$, $3$ and $10$).  In this case, we have used $\epsilon=0.5$ for
the excursion set calculation.

In Sasaki approximation, the destruction rate is independent of mass
and hence the survival probability is also independent of mass.  Our
calculations show that this approximation is not true, and hence the
survival probability of haloes must also depend on mass.  We note that
the survival probability is high for large mass haloes: if a very large
mass halo forms at a high redshift then it is likely to survive
without a significant addition to its mass.  Smaller haloes are highly
likely to merge or accrete enough mass and hence do not survive for
long periods.  Survival probability drops very rapidly as we go to
smaller masses.  While this is expected on physical grounds, it is an
aspect not captured by the Sasaki approximation where equal survival
probability is assigned to haloes of all masses.  The mass dependence
of survival probability is qualitatively similar to that obtained by
\citet{1996MNRAS.280..638K}.  There is no significant qualitative
difference between the curves for the Press-Schechter and the
Sheth-Tormen mass functions.

\subsection{Formation time distribution} 

Finally, another interesting quantity is the distribution
$F(M;t_{f},t)$ of formation epochs $t_{f}$ of haloes with mass $M$ at
$t$, defined in Section \ref{sec:haloform}.  This distribution can be
obtained once the survival probability and formation rate of haloes is
known.  We calculated the formation time distribution using the
excursion set formalism and the Sasaki prescription.  The results are
shown in Fig.~\ref{fig:formation_epoch_distribution}.  We plot
$F(M;z_{f},z=0)/N(M,z=0)$ versus the formation redshift $z_f$ for
three different masses ($10^{13}$, $10^{10}$ and $10^{7} M_{\odot}$)
in the standard $\Lambda$CDM model for both Press-Schechter (left
column) and Sheth-Tormen (right column) mass functions with
$\epsilon=0.5$ (first row) and $\epsilon=0.1$ (second row).  A common
feature is that $F$ as a function of $z_f$ increases up to a certain
redshift and then starts to decline.  The epochs at which $F$ drops by
an order of magnitude from its peak can be interpreted as typical
range of redshifts for the formation of bound systems of respective
masses which exist at $z = 0$.

The differences between the formation redshift distribution for
$\epsilon=0.5$ and $\epsilon=0.1$ are along expected lines: the
formation redshifts are smaller for the lower value of $\epsilon$ as a
smaller change in mass is required for us to declare that a new halo
has formed and hence typical haloes do not survive for a very long
time.  We see that the excursion set calculation suggests that haloes
formed more recently as compared to the Sasaki approximation based
estimate.  This can be understood in terms of the equal survival
probability assigned by the Sasaki approximation to haloes of all
masses.  For a clearer comparison, the ratio of the estimate based on
Sasaki approximation and the excursion set calculation is shown in
Fig.~\ref{fig:formation_epoch_distribution_ratio}.  We note that for
very low mass haloes these two estimates differ by more than an order
of magnitude.  The main qualitative difference between the plots for
the Press-Schechter and the Sheth-Tormen mass functions is caused by
the negative formation rates in the Sasaki approximation.

\subsection{Discussion} 

The results described above show conclusively that the excursion set
approach predicts halo formation and destruction rates that match with
simulations much better than the Sasaki approximation.
 
Another noteworthy aspect is that the destruction and formation rates
depend on the value of $\epsilon$ in simulations as well as the
excursion set calculation thereby allowing us to differentiate between
major and minor mergers.  In comparison, there is no natural way to
bring in this dependence in the Sasaki approximation.  While the match
between simulations and the excursion set approach for the two values
of $\epsilon$ is satisfying, it raises the question of the appropriate
value of this parameter.  In our view the appropriate value of the
parameter should depend on the application in hand.  In semi-analytic
galaxy formation models, we should use a value of $\epsilon$ that
corresponds to the smallest ratio of masses of the infalling galaxy
and the host galaxy where we expect a significant dynamical influence
on star formation rate.  For instance, \citet{1999MNRAS.303..188K} use
$\epsilon=0.3$ in their semi-analytic galaxy formation model while
considering formation of bulges in merger remnants.  In case of galaxy
clusters we may base this on the smallest ratio of masses where the
intra-cluster medium is likely to be disturbed in a manner accessible
to observations in X-ray emission or the Sunyaev-Zel'dovich effect
\citep{1972CoASP...4..173S, 1995MNRAS.275..720N, 2004MNRAS.347L..13K}.

While the close match between simulations and the excursion set
calculation is useful, it also implies that we should not use the
simpler Sasaki approximation.  The excursion set calculation of the
halo destruction rate is fairly simple for the Press-Schechter mass
function, but the corresponding calculation for the Sheth-Tormen mass
function is much more complicated.  Plots of the destruction rate
efficiency $\phi(M)$ for all the models suggest that its variation
with mass and $\epsilon$ is very similar for the PS and ST mass
function.  This suggests an approximation where we use
$\phi(M,z;\epsilon)$ computed using the Press-Schechter mass function
and use that to compute the halo formation rate in the Sheth-Tormen
mass function.  Figs.~\ref{fig:lcdm_approx_eps0.5} and
\ref{fig:lcdm_approx_eps0.1} show the halo formation rate for the
$\Lambda$CDM model at different redshifts and compare the excursion set
calculation, the Sasaki approximation and the intermediate
approximation suggested above. We have also shown the ratios of formation 
rates estimated in these two approaches mentioned above in 
Figs.~\ref{fig:lcdm_approx_eps0.5_ratio} and \ref{fig:lcdm_approx_eps0.1_ratio}. 
We find that the intermediate 
approximation is not plagued by negative halo formation rates and that
it is an excellent approximation at all mass scales at higher
redshifts.  At lower redshifts, the approximation is still good at
high masses but not so at smaller masses. 

While comparing our analytical results with those of N-body
simulations, we find a systematic deviation between the two at the
high mass end.  This is possibly related to the problm of `halo
fragmentation' while deriving halo merger trees from the simulations.
In about 5\% of all haloes, particles in a given progenitor halo can
become part of two independent haloes at a future epoch.  This is
usually attributed to the fact that the FOF algorithm groups particles
based on the inter-particle distance.  This can result in the
identification of two haloes separated by a thin 'bridge' of particles
to be treated as a single halo.  Such halo fragmentation has been
treated using different techniques in various halo formation rate
studies. \citet{2008MNRAS.386..577F} compare these techniques and find
that the effect of halo fragmentation is maximum of high mass haloes.

\section{Conclusions}

Key points presented in this paper can be summarized as follows:
\begin{itemize}
\item
We revisit the Sasaki approximation for computing the halo formation rate and
compute the destruction rate explicitly using the excursion set approach.
\item
We introduce a parameter $\epsilon$, the smallest fractional change in mass of
a halo before we consider it as destruction of the old halo and formation of a
new halo.  
\item
We show that the halo destruction rate is not independent of mass even for
power law models and hence the basis for the Sasaki ansatz does not hold.
Two prominent features of the halo destruction rate are the rapid fall at
large masses, and a pronounced peak close to the scale of non-linearity.  The
peak is more pronounced for smaller values of $\epsilon$.
\item
Using the excursion set approach for the Sheth-Tormen mass function leads to
positive halo formation rates, unlike the generalization of the Sasaki ansatz
where formation rates at some mass scales are negative.
\item
We compare the destruction rate and the halo formation rates computed using
the excursion set approach with N-body simulations.
We find that our approach matches well with simulations for all models,
at all redshifts and also for different values of $\epsilon$.
\item
In some cases there are deviations between the simulations and the theoretical
estimate.  
However, these deviations are much smaller for the excursion set based method
as compared to the Sasaki estimate. 
\item
It may be that some of the deviations arise due to a series representation of
the barrier shape, and the number of terms taken into account may not suffice
for the estimate.  
We have found that truncation of the series can affect results at small
masses, though in most cases results converge with the five terms that we have
taken into account for the range of masses considered here.
\item
We show that we can use the halo destruction rate computed for the
Press-Schechter mass function to make an approximate estimate of the
halo formation rate in Sheth-Tormen mass function using
equation (\ref{eq:Ndotform2}). 
This approximate estimate is fairly accurate at all mass scales in the
$\Lambda$CDM model at high redshifts.
\item
The halo survival probability is a strong function of mass of haloes, unlike
the mass independent survival probability obtained in the Sasaki
approximation. 
\item
The halo formation redshift distribution for haloes of different masses is also
very different from that obtained using the Sasaki approximation. 
This is especially true for the Sheth-Tormen mass function where the Sasaki
approximation gives negative halo formation rates in some range of mass scales
and redshifts.
\end{itemize}

The formalism used here for calculation of halo formation rate and other
related quantities can be generalized to any description of the mass function
if the relevant probabilities can be calculated. 
Within the framework of the universal approach to mass functions, it can also
be used to study formation rates of haloes in different cosmological models
\citep{2003MNRAS.346..573L, 2004PhRvD..69l3516M}. 
This allows for an easy comparison of theory with observations for quantities
like the major merger rate for galaxy clusters \citep{2001MNRAS.325.1053C}.

In case of semi-analytic models of galaxy formation, our approach allows for a
nuanced treatment where every merger need not be treated as a major merger and
we may only consider instances where mass ratios are larger than a critical
value for any affect on star formation in the central galaxy. 


\section*{Acknowledgments}

Computational work for this study was carried out at the cluster
computing facility in the Harish-Chandra Research Institute
(http://cluster.hri.res.in/index.html).
JSB thanks Ravi Sheth and K Subramanian for useful discussions. 
This research has made use of NASA's Astrophysics Data System.


\label{lastpage}

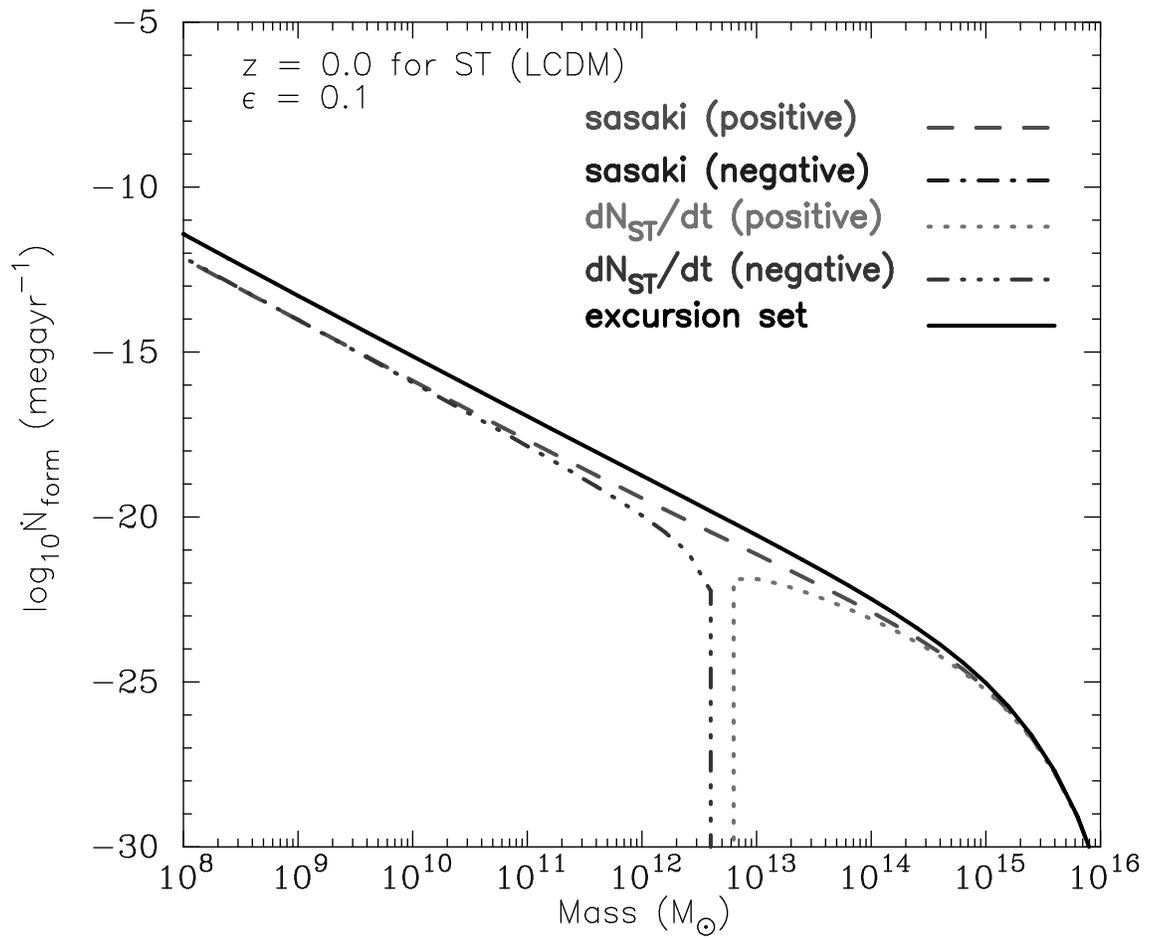
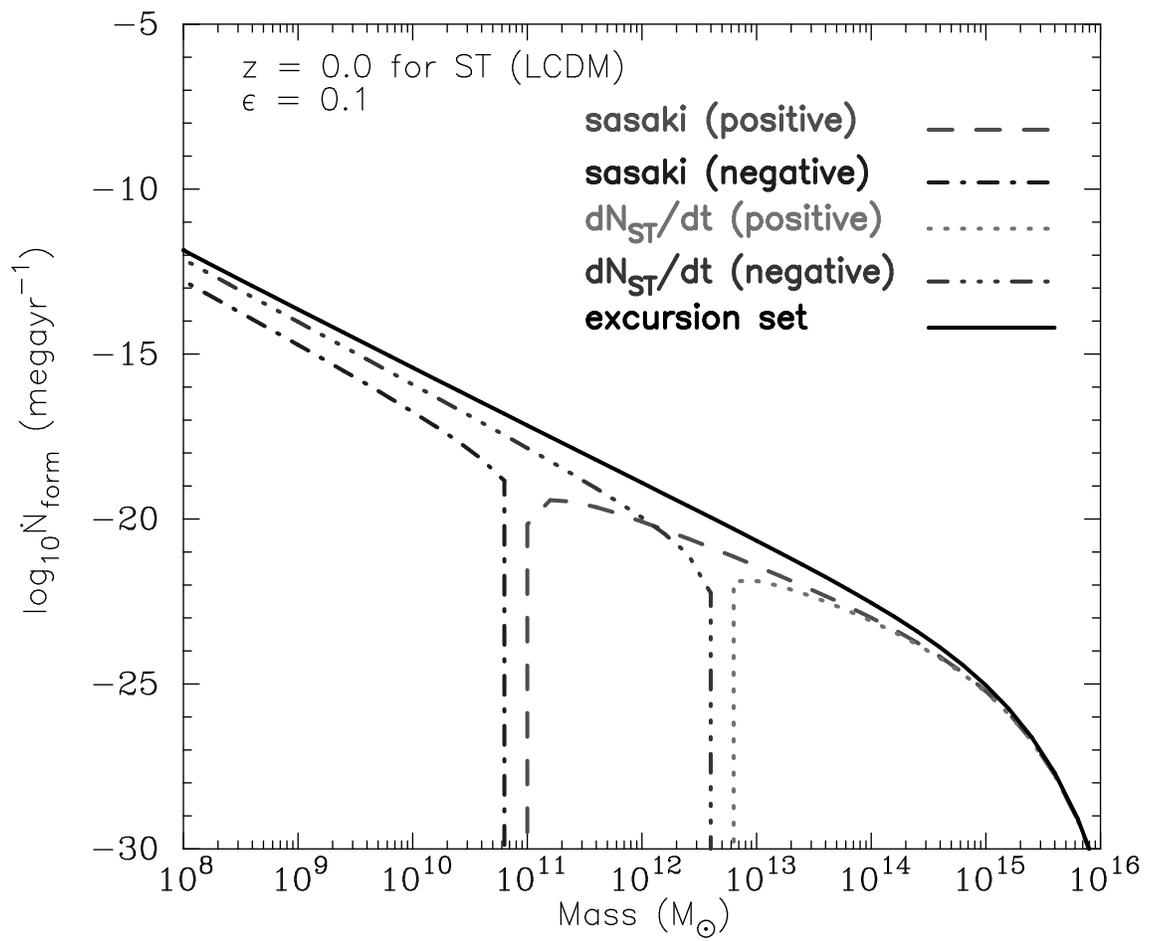
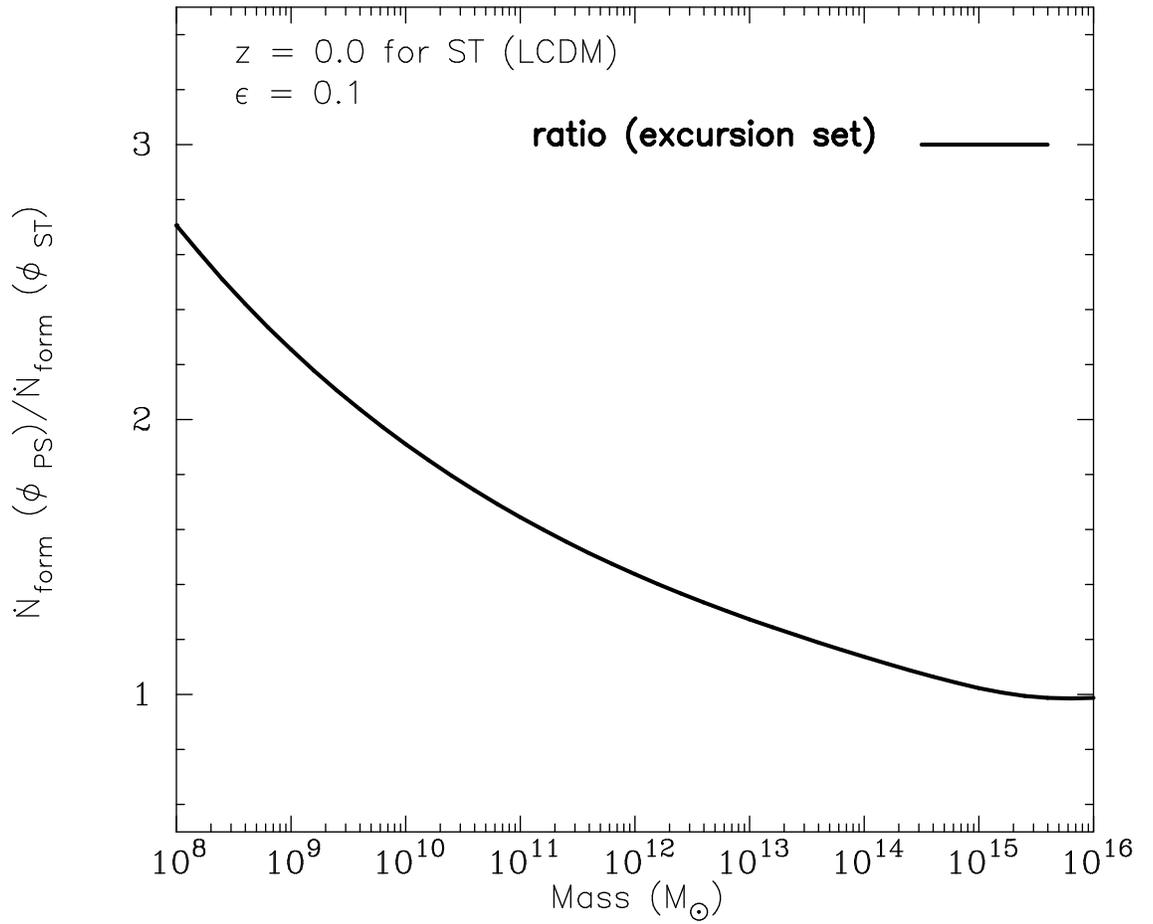
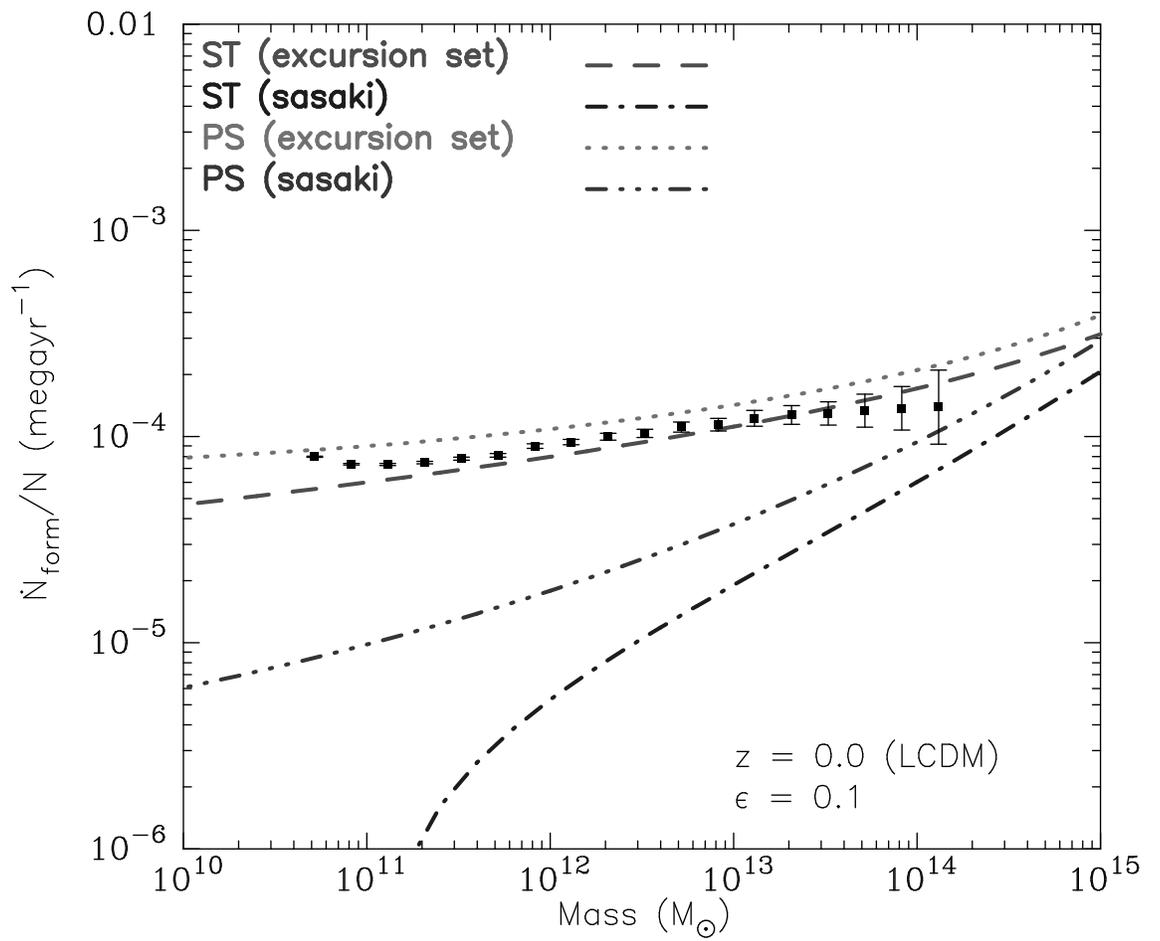
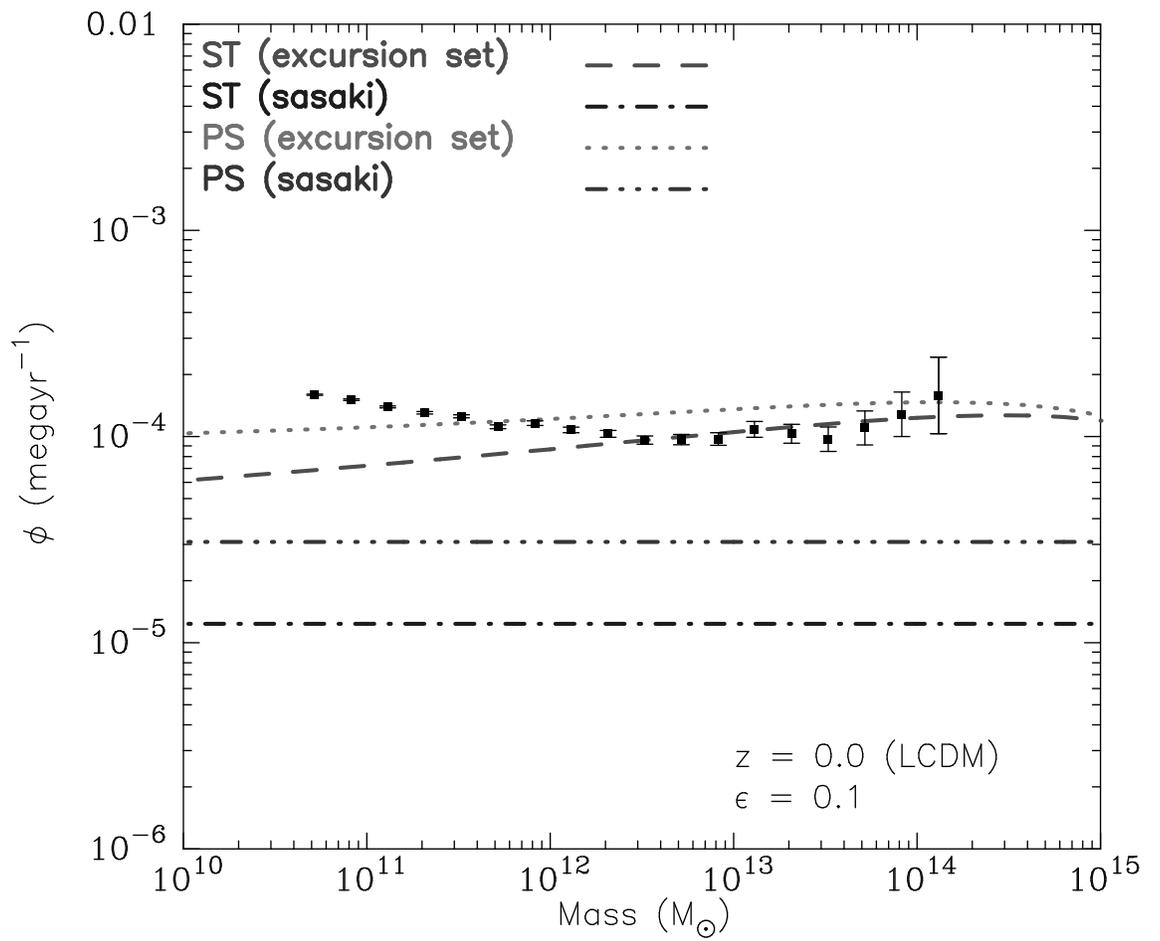
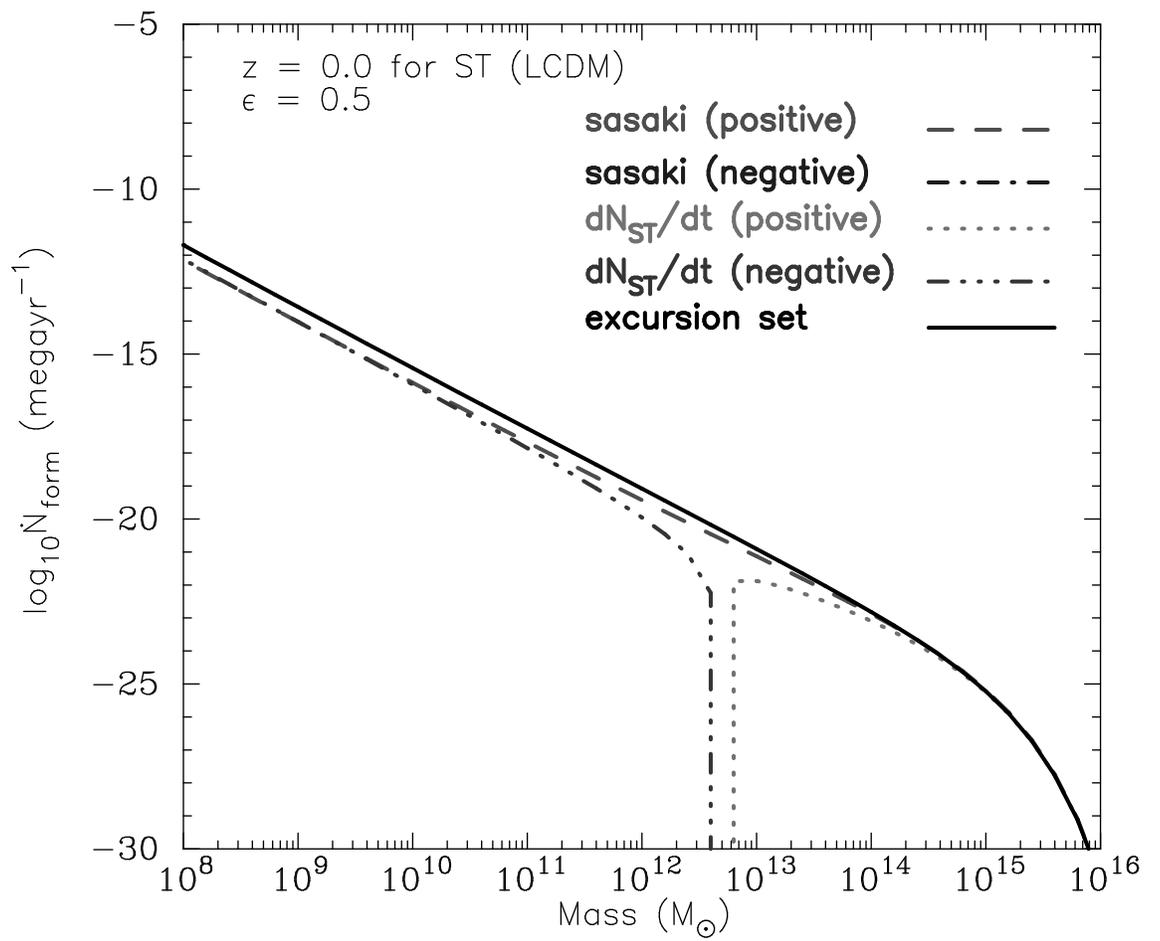
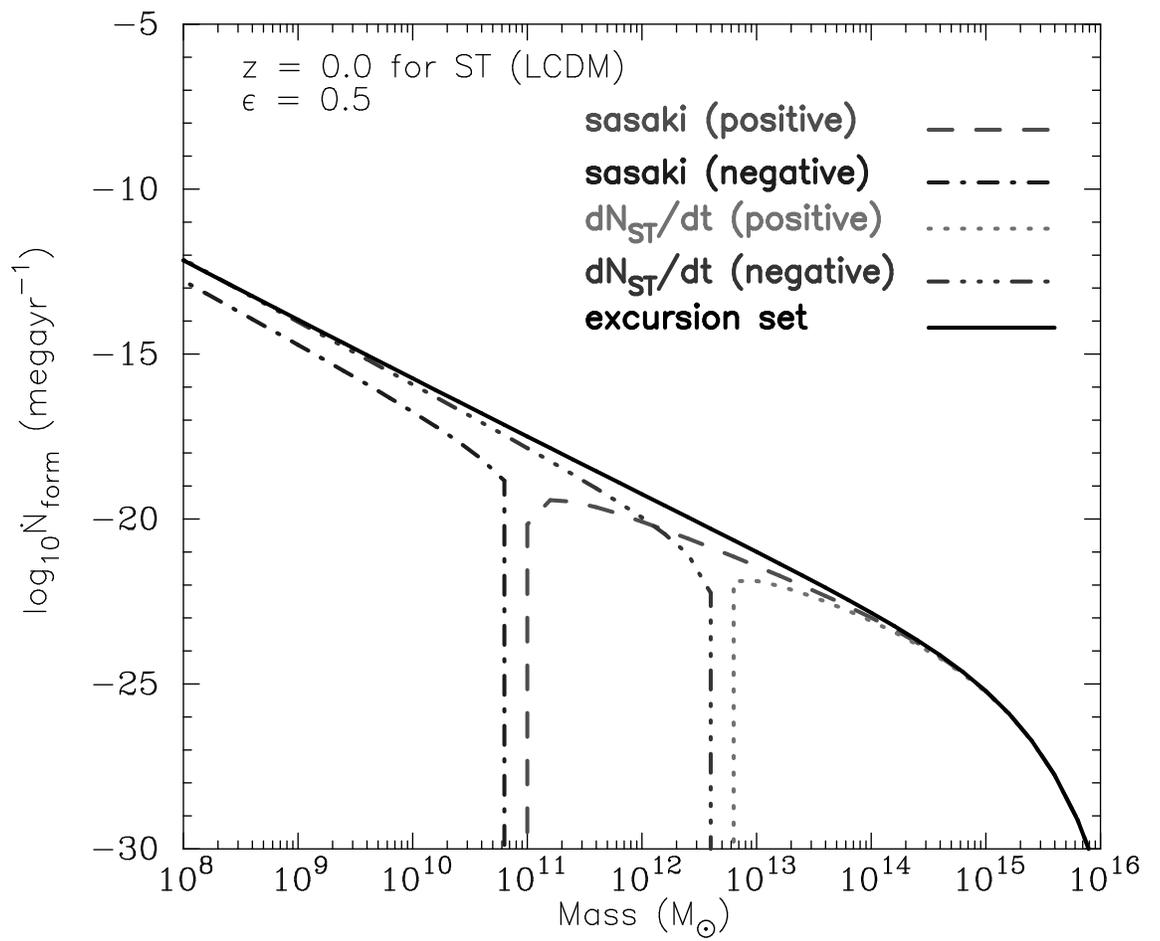
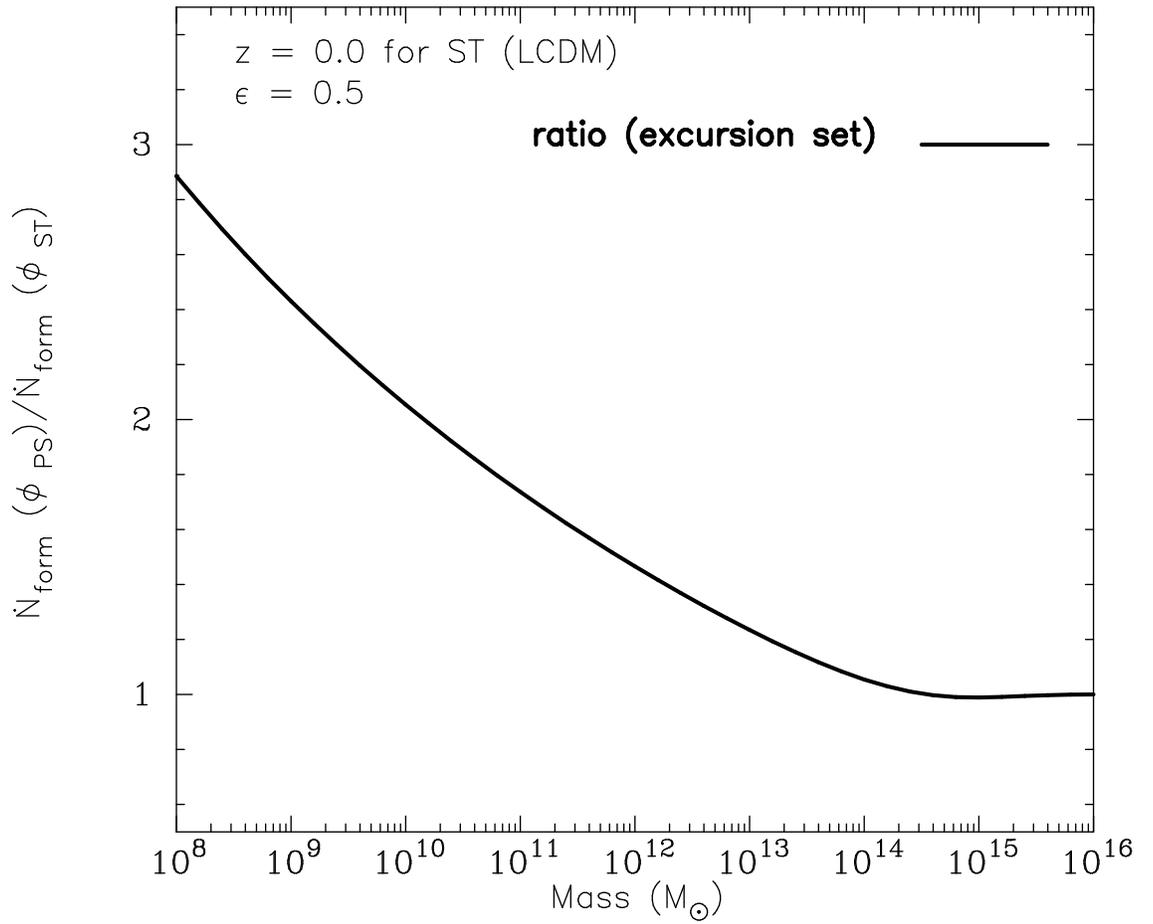
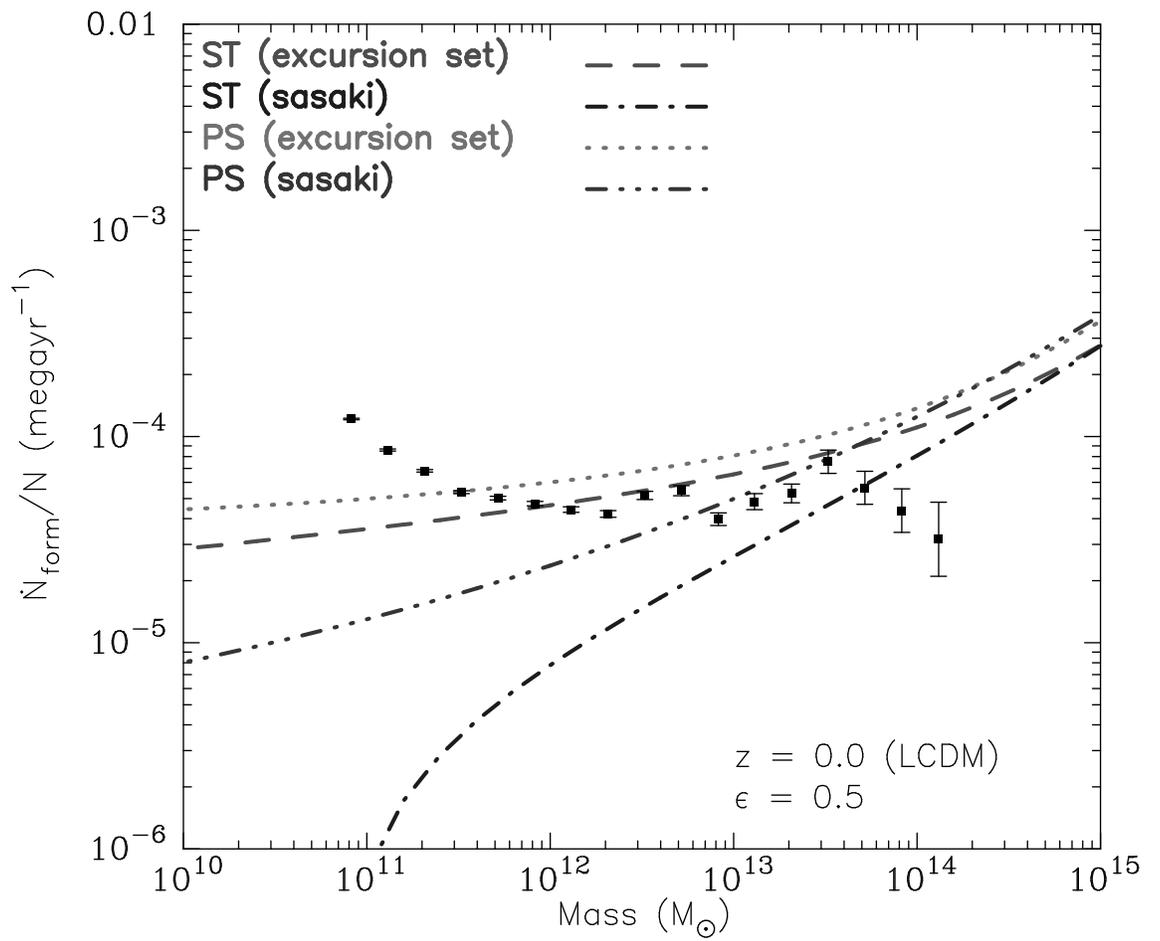
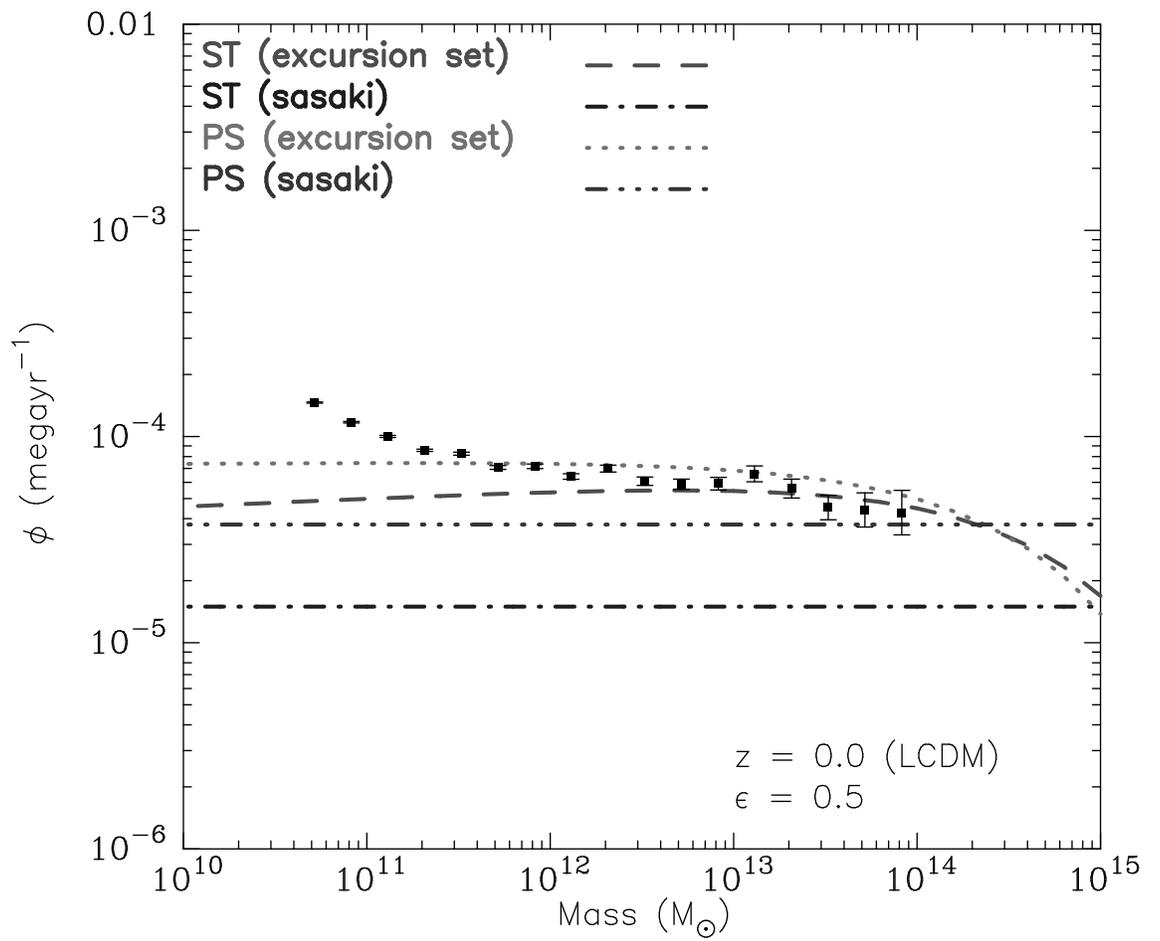
\end{document}